\documentclass[12pt]{nature}
\usepackage{lineno,soul}

\usepackage{graphicx}
\usepackage{amssymb, amsmath}
\usepackage{color}
\usepackage[dvipsnames]{xcolor}
\usepackage{hyperref}
\makeatletter
\let\saved@includegraphics\includegraphics
\AtBeginDocument{\let\includegraphics\saved@includegraphics}
\renewenvironment*{figure}{\@float{figure}}{\end@float}
\makeatother

\graphicspath{{./}{figs/}}

\newcommand{\rev}[1]{{#1}} 
\newcommand{\frev}[1]{{#1}} 

\usepackage[T1]{fontenc} 
\usepackage[utf8]{inputenc} 

\usepackage{bibunits}
\usepackage{nature}

\title{\rev{Gravitational instability in a planet-forming disk} }
\author{J. Speedie$^{1}$, R. Dong$^{1,2}$, C. Hall$^{3,4}$, C. Longarini$^{5,6}$, B. Veronesi$^{7}$, T. Paneque-Carreño$^{8,9}$, G. Lodato$^{5}$, Y. Tang$^{10}$, R. Teague$^{11}$, J. Hashimoto$^{12,13,14}$}

\begin{document}

\maketitle

\begin{affiliations}
 \item Department of Physics \& Astronomy, University of Victoria, Victoria, BC, V8P 5C2, Canada 
 \item Kavli Institute for Astronomy and Astrophysics, Peking University, Beijing 100871, People’s Republic of China 
 \item Department of Physics and Astronomy, The University of Georgia, Athens, GA 30602, USA 
\item Center for Simulational Physics, The University of Georgia, Athens, GA 30602, USA 
 \item Università degli Studi di Milano, Via Celoria 16, 20133, Milano, Italy 
 \item Institute of Astronomy, University of Cambridge, Madingley Road, Cambridge, CB3 0HA, United Kingdom 
 \item Univ Lyon, Univ Lyon1, Ens de Lyon, CNRS, Centre de Recherche Astrophysique de Lyon UMR5574, F-69230, Saint-Genis,-Laval, France 
 \item Leiden Observatory, Leiden University, P.O. Box 9513, NL-2300 RA Leiden, the Netherlands 
 \item European Southern Observatory, Karl-Schwarzschild-Str 2, 85748 Garching, Germany 
 \item Academia Sinica, Institute of Astronomy and Astrophysics, 11F of AS/NTU Astronomy-Mathematics Building, No.1, Sec. 4, Roosevelt Rd., Taipei, Taiwan 
 \item Department of Earth, Atmospheric, and Planetary Sciences, Massachusetts Institute of Technology, Cambridge, MA 02139, USA 
 \item Astrobiology Center, National Institutes of Natural Sciences, 2-21-1 Osawa, Mitaka, Tokyo 181-8588, Japan 
 \item Subaru Telescope, National Astronomical Observatory of Japan, Mitaka, Tokyo 181-8588, Japan 
 \item Department of Astronomy, School of Science, Graduate University for Advanced Studies (SOKENDAI), Mitaka, Tokyo 181-8588, Japan 
\end{affiliations}

\small

\begin{bibunit}[sn-nature]

\begin{abstract}
\rev{
The canonical theory for planet formation in circumstellar disks proposes that planets are grown from initially much smaller seeds\cite{chiang-youdin-2010-review, johansen-lambrechts2017-review, ormel2017-review, liu-ji-2020-review, drazkowska2023-ppvii}. 
The long-considered alternative theory 
proposes that  
giant protoplanets can be formed directly from collapsing fragments of vast spiral arms\cite{boss1997, gammie2001, rice2003, zhu2012-challenges-clumps, deng2021-magnetic-fragmentation, cadman2021} induced by gravitational instability (GI)\cite{lodato-rice-2004, cossins2009, dipierro2014} -- \textit{if} the disk is gravitationally unstable.
For this to be possible, the disk must be massive compared to the central star: a disk-to-star mass ratio of $1/10$ is widely held as the rough threshold for triggering GI, \frev{inciting significant non-Keplerian dynamics} and generating prominent spiral arms\cite{kratter-lodato-2016, dong2015-GIspirals-scatteredlight, hall2016-continuumGIspirals, hall2019-temporalGIspiralsALMA}. 
While estimating disk masses has historically been challenging\cite{panequecarreno2021-elias27, veronesi2021-elias27, stapper2023-herbig-gas-masses}, the motion of the gas can reveal the presence of GI through its effect on the disk velocity structure\cite{hall2020, longarini2021, terry2022-diskmass}.} 
Here we present kinematic evidence \rev{of} gravitational instability in the disk around AB Aurigae,  
using deep observations of $^{13}$CO and C$^{18}$O line emission 
with the Atacama Large Millimeter/submillimeter Array (ALMA). 
The observed kinematic signals strongly resemble  
predictions from simulations 
and analytic \rev{modeling}. 
From quantitative comparisons, we infer \frev{a disk mass of up to $1/3$ the stellar mass enclosed within 
$1''$ to $5''$ on the sky.} 
\end{abstract}


We targeted the disk around AB Aurigae (AB Aur), a 
$2.5-\rev{4.4}$ Myr old\cite{vandenancker1997, dewarf2003, beck2019-H2, garufi2024-destinys-taurus} 
Herbig Ae\cite{rodrigues2014-abaur-outflow} 
star of intermediate mass ($M_{\star} = 2.4 \, M_\odot$)\cite{dewarf2003, beck2019-H2, guzman-diaz2021-herbig-study} 
at a distance of $155.9 \pm 0.9$ pc\cite{gaiaDR3-2023}.  
AB Aur is at a relatively late stage of 
protostellar evolution,  
classified as a Class II Young Stellar Object\cite{henning1998, bouwman2000} (YSO). 
To probe the velocity structure of the disk, 
we obtained deep ALMA Band 6 observations of \frev{molecular} emission lines 
$^{13}$CO ($J=2-1$) and C$^{18}$O ($J=2-1$) 
with high velocity resolution (channel widths of $v_{\rm chan}=42$ m/s and $84$ m/s respectively). 
The observations were taken in two array configurations with baselines ranging $14$ to $2,216$ m,  
reaching a total 
on-source integration time 
of 5.75 hours.
Imaging with a Briggs robust value of $0.5$ provided image cubes with 
a spatial resolution or beam size of $0.237'' \times 0.175''$ (beam position angle, ${\rm PA}=1.2^\circ$) equivalent to $37 \times 27$ au. 
\rev{We collapse the 3D image cubes into 2D moment maps to expose the velocity-integrated intensity (moment 0), intensity-weighted line-of-sight velocity ($v_{\rm los}$, moment 1) and emission line width (moment 2). This collection is} shown in Extended Data Figure~\ref{extfig:momentmaps}.

\begin{figure}
\centering
\includegraphics[width=0.9\textwidth,angle=0]{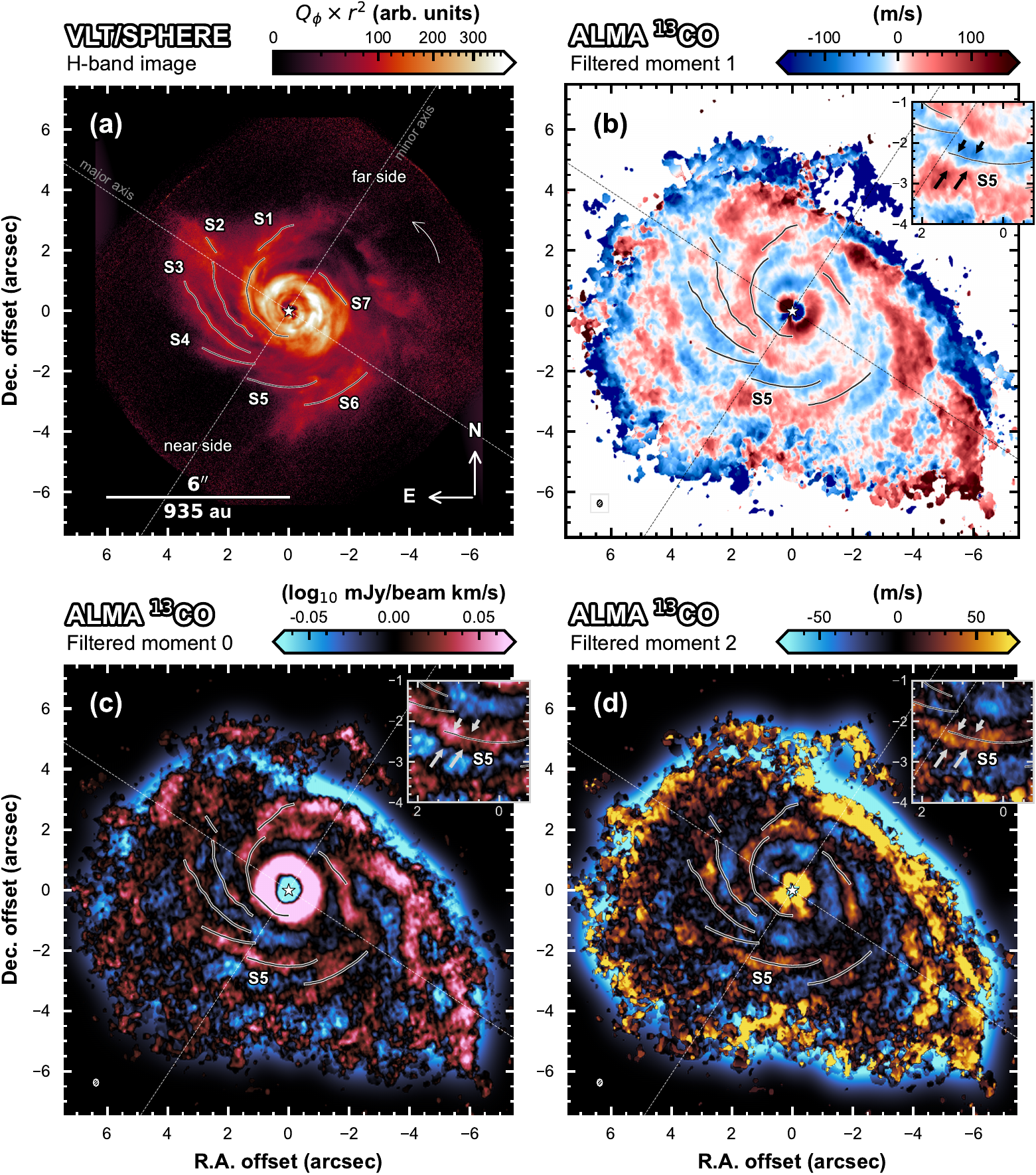}
\caption{{\bf Global spirals in the AB Aur disk.} 
\textbf{(a)} VLT/SPHERE $H$-band scattered light image of the AB Aur disk (ref.\cite{boccaletti2020-abaursphere}) 
tracing spiral structure in (sub-)micron-sized dust grains.  
The labelled spirals S1-S7 are taken from previous works (ref.\cite{fukagawa2004, hashimoto2011}). 
\textbf{(b)} Filtered ALMA $^{13}$CO intensity-weighted mean velocity (moment 1) map, revealing residual gas motion within the bulk flow. \rev{The synthesised beam is shown in the bottom left corner as an ellipse.} 
The inset zooms into the region around where S5 crosses the minor axis, highlighting converging flows on the two sides of S5 indicated by arrows. 
\textbf{(c)} Filtered ALMA $^{13}$CO integrated intensity (moment 0) map, highlighting peaks in the gas density and/or temperature.
\textbf{(d)} Filtered ALMA $^{13}$CO emission line width (moment 2) map, showing localized line broadening within the spiral arms. 
\rev{Insets in panels (c,d) zoom into the same region as panel (b) inset, showing enhanced gas density/temperature} caused by the radially converging flows around S5.
}
\label{fig:1:residuals}
\end{figure}

To reveal the 
spiral arms in the disk, we apply a high-pass filter\cite{perez2016-elias27} (see Methods) to the ALMA $^{13}$CO moment maps (Figure~\ref{fig:1:residuals}bcd). 
\rev{In the} filtered 
line-of-sight velocity (moment 1) map, we observe spiral-shaped disturbances in the gas velocity field throughout the disk (Figure~\ref{fig:1:residuals}b).
With the filtered velocity-integrated intensity (moment 0) and line width (moment 2) maps, we visually highlight regions of peak density and temperature (Figure~\ref{fig:1:residuals}cd). 
Compression and shock-heating are expected to lead to temperature enhancements (and thus localized line broadening) within GI-induced density spirals in self-regulating disks\cite{cossins2009, longarini2021}. 
The VLT/SPHERE $H$-band scattered light image of AB Aur originally presented in Boccaletti et al. (2020)\cite{boccaletti2020-abaursphere}
is shown for comparison (Figure~\ref{fig:1:residuals}a). 
Scattered light comes from the disk surface, probing the distribution of (sub-)micron-sized dust usually well-coupled with the gas. Previous simulations have shown that GI-induced density spirals are prominent in scattered light\cite{dong2015-GIspirals-scatteredlight, dong16protostellar}.  
At least seven spiral structures (S1-S7) have been previously identified in the $H$-band image\cite{hashimoto2011, boccaletti2020-abaursphere}, 
though not all occupy the same radial region and 
some may be branches of adjacent arms\cite{fukagawa2004}. 
\rev{The disk rotates counter-clockwise (the spiral arms are trailing), and the south side is the near side, tilted toward us\cite{fukagawa2004, lin2006-possible-molecular-spiralarms, perrin2009}.} 

To provide a qualitative comparison to the ALMA observations, we run 3D smoothed-particle hydrodynamic (SPH) simulations of a gravitationally unstable disk (see Methods). 
The simulations were post-processed with radiative transfer and then further processed to have the same viewing angle, sensitivity, spectral and angular resolution as the AB Aur data. 
To place the disk comfortably within the gravitationally unstable regime \rev{($M_{\rm disk}/M_{\star} \gtrsim 0.1$)}, we set the total gas mass to \rev{$0.3\times$} the mass of the star. 
\rev{For sustained spiral arms, we set} the cooling timescale to $10\times$ the local dynamical timescale ($\beta = 10$). 
The simulated GI disk shows spiral structures in all three moment maps, resembling those in the AB Aur disk 
(Extended Data Figures~\ref{extfig:momentmaps} and \ref{extfig:momentmaps-filtered}). 
Overall, the AB Aur disk hosts a global architecture 
of spiral arms at 100 to 1,000 au scales across all azimuths in multi-wavelength observations tracing different disk components and quantities, strongly indicating ongoing gravitational instability. 

\begin{figure}
\includegraphics[width=\textwidth,angle=0]{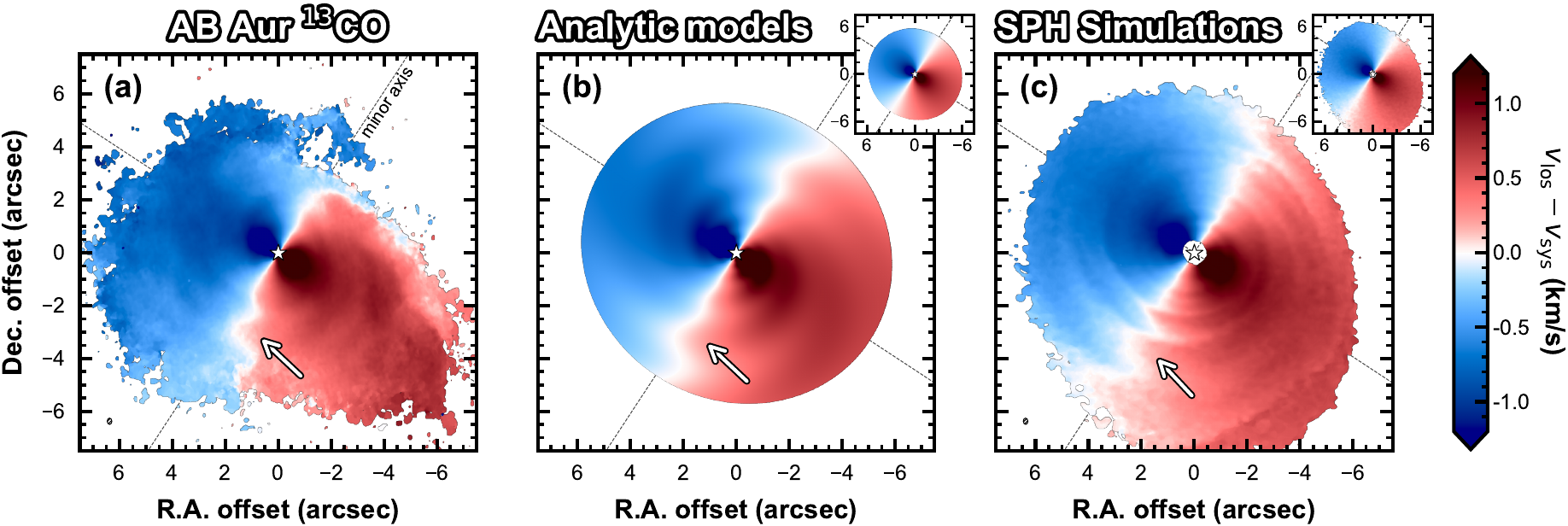}
\caption{\rev{{\bf Detection of the GI wiggle 
in the AB Aur disk.}}
\textbf{(a)} ALMA $^{13}$CO intensity-weighted mean velocity 
(moment 1) map showing line-of-sight velocity ($v_{\rm los}$) of gas in the AB Aur disk. The observations display the ``GI wiggle'' along the minor axis (arrow) predicted by Hall et al. (2020)\cite{hall2020} as a clear kinematic signature of gravitational instability.  
\textbf{(b)} $v_{\rm los}$ map of a gravitationally unstable disk at the inclination and position angle of the AB Aur disk, computed with 2D analytic \rev{modeling}\cite{longarini2021}. Self-gravitating spiral arms crossing the minor axis induce radial motion that appears as a wiggle (arrow). 
\textbf{(c)} Synthetic \rev{ALMA $^{13}$CO} moment 1 map of the 3D SPH GI disk simulation, revealing the same GI wiggle signature (arrow).
The insets in panels (b) and (c) show corresponding images for
Keplerian disks with no radial gas motion, \rev{where} 
the isovelocity curve at the systematic velocity appears as a straight line along the minor axis.
 }
\label{fig:2:pp-wiggle}
\end{figure}

One characteristic kinematic feature in the AB Aur disk can be found in the isovelocity curve at the systemic velocity $v_{\rm sys}$
in the moment 1 map --- 
Figure~\ref{fig:2:pp-wiggle}a shows 
a sinusoidal pattern at $v_{\rm los}=v_{\rm sys}$ (along the minor axis; white color), more prominent towards the south. 
This signature, known as a ``minor axis GI wiggle''\cite{hall2020}, 
has been predicted in hydrodynamic simulations\cite{hall2020, terry2022-diskmass} and analytic theory\cite{longarini2021} as a clear kinematic signature of gravitational instability (Figure~\ref{fig:2:pp-wiggle}bc). 
It is one of a global set of GI wiggles in isovelocity curves 
we observe 
throughout the AB Aur disk (Extended Data Figure~\ref{extfig:global-wiggles}). 
These wiggles are generated by self-gravitating spiral arms, which constitute local minima in the gravitational potential field and induce corresponding oscillations in the gas velocity field. 
The 
synthetic moment 1 map of the SPH GI disk simulation shows 
a minor axis GI wiggle 
with similar morphology as the observed one (Figure~\ref{fig:2:pp-wiggle}c),  
completely distinct from the linear pattern found in 
a disk undergoing Keplerian rotation with no radial motions (Figure~\ref{fig:2:pp-wiggle}bc insets). 

Among all GI wiggles, the minor axis GI wiggle has been known and targeted in past studies for its convenience in quantitative analysis\cite{longarini2021, terry2022-diskmass}.
Due to projection effects, only the radial and vertical components of the disk velocity field ($v_{r}$ or $v_{z}$) 
contribute to $v_{\rm los}$ at the systemic velocity traced by this wiggle. 
In the case of GI-induced velocity perturbations, the $v_{r}$ contribution is expected to dominate\cite{hall2020}. 
As we show with 2D analytic calculations of gravitationally unstable disks (see Methods), 
a self-gravitating spiral arm induces radial motion convergent on itself, appearing as a wiggle in the moment 1 map at $v_{\rm sys}$
where the spiral crosses the minor axis (c.f. \rev{Extended Data Figure~\ref{extfig:kinematics-spirals}}). 
The filtered moment 1 map in Figure \ref{fig:1:residuals}b displays red\rev{-} and blue\rev{-shift} patterns corresponding to convergent flows toward spiral S5 (visible in both scattered light and $^{13}$CO moment 0 and 2; Figure~\ref{fig:1:residuals}acd), supporting the interpretation that the GI wiggle along the southern minor axis in Figure \ref{fig:2:pp-wiggle}a is generated by a self-gravitating spiral arm.

\begin{figure}
\includegraphics[width=\textwidth,angle=0]{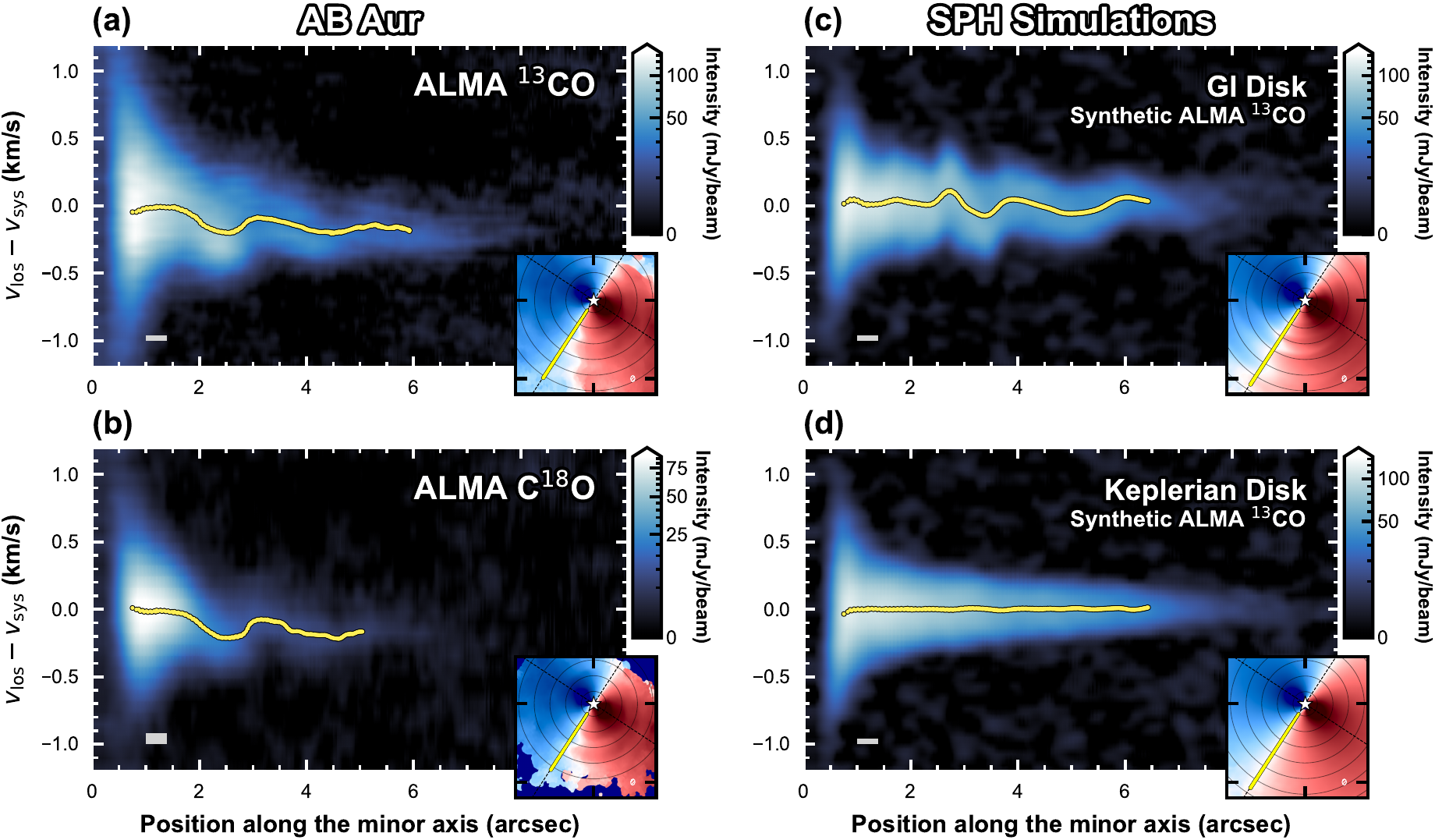}
\caption{\rev{{\bf The GI wiggle in position-velocity space, or PV wiggle.}} 
Emission spectra (intensity as a function of velocity) extracted along the southern minor axis of the disk, plotted with distance from the star. The line centers 
are shown as yellow points.  
\rev{The insets at the bottom right of each panel show the corresponding line center map, with black circles delineating $1''$ radial increments. The yellow line along the southern minor axis is the narrow ($0.5^{\circ}$-wide) wedge-shaped mask within
which the spectra and line centers are extracted.} 
In all \rev{PV diagram} panels, the grey box in the bottom left corner has horizontal width equal to the beam major axis and vertical height equal to the channel width. 
\rev{\textbf{(a)} ALMA observations of the AB Aur disk in $^{13}$CO and \textbf{(b)} in C$^{18}$O. 
\textbf{(c,d)} Synthetic ALMA $^{13}$CO observations   
generated from 3D 
SPH simulations of a gravitationally unstable disk with a 
disk-to-star mass ratio of 0.3 and a cooling rate described by $\beta=10$. 
In \textbf{(d)}, the simulated disk has its velocity structure artificially post-processed 
to be Keplerian. 
}
}
\label{fig:3:spectra}
\end{figure}

\begin{figure}
\includegraphics[width=\textwidth,angle=0]{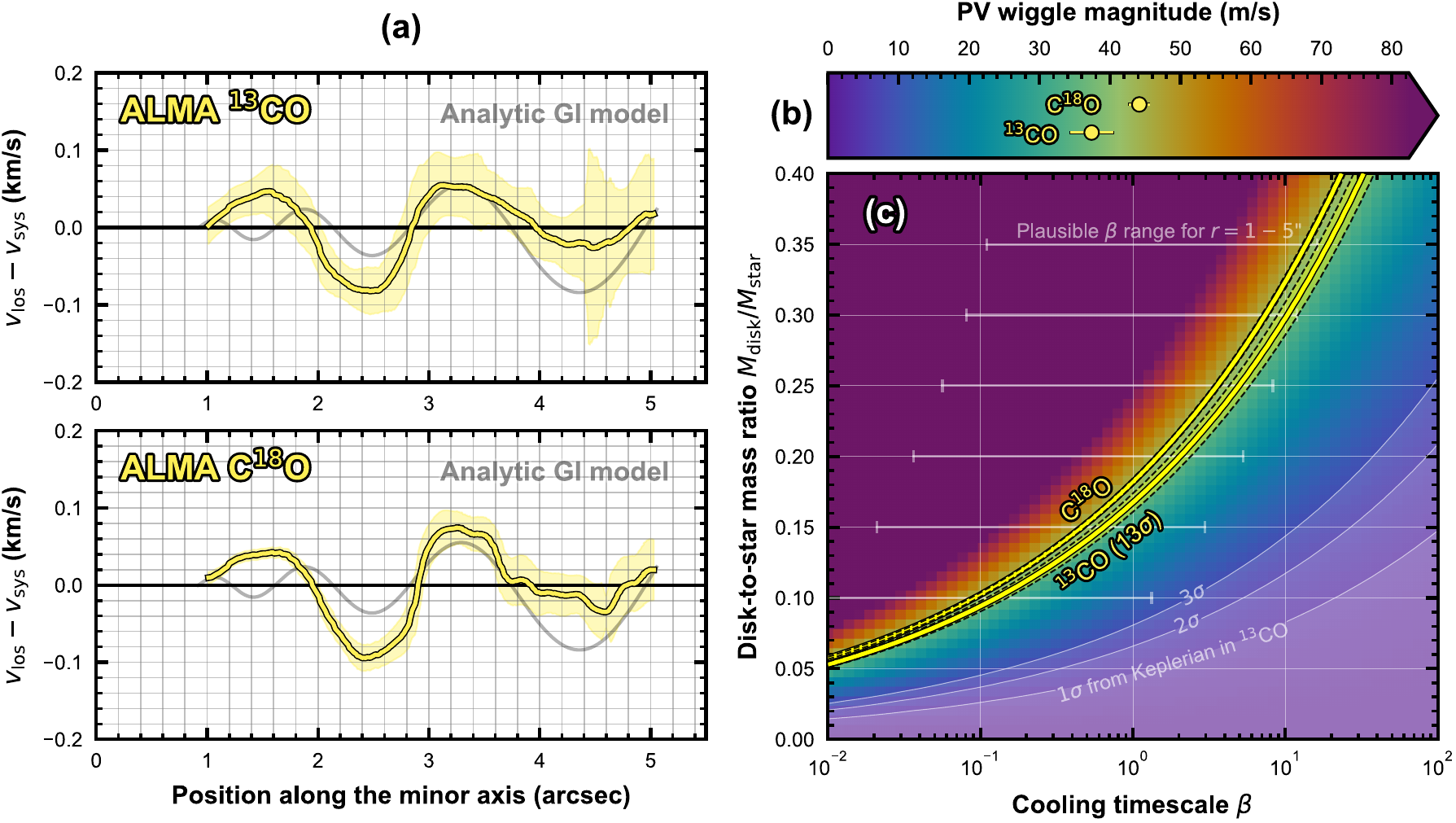}
\caption{\rev{{\bf PV wiggle mophology, magnitude, and constraints on the AB Aur disk mass.}} 
\textbf{(a)} \rev{The ALMA $^{13}$CO and C$^{18}$O line centers along the southern minor axis from \rev{Figure~\ref{fig:3:spectra}ab}, after quadratic detrending (Methods). 
Uncertainties on the line centers are shown by yellow shaded regions.} 
\rev{For qualitative comparison, the minor axis PV wiggle of the analytic GI model disk from Figure~\ref{fig:2:pp-wiggle}b is shown in the background in light grey.}  
\textbf{(b)} The magnitude of the southern minor axis PV wiggle in AB Aur is measured to be \rev{$37.4 \pm 2.9$ m/s in $^{13}$CO and $44.2 \pm 1.3$ m/s in C$^{18}$O.}  
\textbf{(c)} A map of the minor axis PV wiggle magnitude of 3600 analytic GI model disks, calculated for a $60\times60$ grid of disk-to-star mass ratios and cooling timescales. Each cell in the map represents the minor axis PV wiggle magnitude from a different model. 
A \frev{yellow} contour is drawn at each of the AB Aur $^{13}$CO and C$^{18}$O measured magnitude values, and dashed lines represent the quoted uncertainties. 
\frev{White shaded regions denote $1\sigma$, $2\sigma$ and $3\sigma$ departures from a Keplerian signal in $^{13}$CO (c.f. Fig. \ref{fig:3:spectra}d).} 
\frev{White horizontal bars 
indicate independently derived $\beta$ ranges at a selection of $M_{\rm disk}/M_{\star}$ values (Methods).}
}
\label{fig:4:amplitude}
\end{figure}

Having identified evidence of gravitational instability in disk kinematics and in the detections of spirals across multiple tracers and moment maps, we now quantitatively analyze the GI wiggle along the southern minor axis to constrain the disk mass. 
We extract the $^{13}$CO and C$^{18}$O emission spectra along the southern disk minor axis (Figure~\ref{fig:3:spectra}\rev{ab}) and 
detect the wiggle in position-velocity space (hereafter referred to as the ``PV wiggle''), which is a different view of the position-position wiggle in Figure~\ref{fig:2:pp-wiggle}a. Slicing the \rev{3D image} cubes this way more comprehensively exposes the gas velocity structure and enables us to quantify the \rev{perturbation} 
in units of velocity.  
We measure the emission line centers by performing a quadratic fit to the spectrum in each spatial pixel of the image cube\cite{teague2018-robust-linecentroids}. 
This method achieves sub-spectral resolution precision on the line center and yields statistically meaningful and robust uncertainties\cite{teague2019-statistical-uncertainties}. 
We find remarkabl\rev{y} similar \rev{sinusoidal} 
morphology between the PV wiggles in $^{13}$CO and C$^{18}$O emission \rev{(Figure~\ref{fig:4:amplitude}a)}. 

Theoretical studies have shown that the dynamical response of a disk 
to its own self-gravity is sensitive to the disk-to-star mass ratio and the cooling rate\cite{longarini2021, terry2022-diskmass}. 
Specifically, the amplitude of the induced radial velocity perturbations is proportional to $(M_{\rm disk}/M_{\star})^2$ and $\beta^{-1/2}$ (Eqns. \ref{eqn:vr-amplitude} \& \ref{eqn:analytic-PV-wiggle} in Methods). 
This allows us to use the observed minor axis PV wiggle to infer the disk mass once we make assumptions on the disk cooling rates. 
Following Longarini et al. (2021)\cite{longarini2021}, we employ a statistical metric to quantify the `magnitude' of the minor axis PV wiggle,  
defined as the standard deviation of the line center velocities over a radial range. \rev{Bounded by the inner central cavity and outer edge of recovered C$^{18}$O emission, our radial range spans $1''$ to $5''$ ($155$ to $780$ au)}.  
We find a magnitude of \rev{$37.4 \pm 2.9$ m/s} 
for the southern minor axis PV wiggle in $^{13}$CO and 
\rev{$44.2 \pm 1.3$ m/s} 
in C$^{18}$O (Figure~\ref{fig:4:amplitude}b). 
For comparison, the gravitationally unstable disk in the SPH simulation has a southern minor axis PV wiggle in $^{13}$CO emission
with quantitatively similar amplitude and \rev{sinusoidal} morphology (\rev{Figure~\ref{fig:3:spectra}c}), 
and a magnitude of $\rev{39.1} \pm 1.8$ m/s (Extended Data Figure~\ref{extfig:pv-wiggle-amp:sim}\rev{a}).

Quantifying the minor axis PV wiggle magnitude as above, we perform 
comparisons against analytic models to identify the combinations of disk mass ($M_{\rm disk}/M_{\star}$) and cooling timescale ($\beta$) 
that satisfy the AB Aur observations. A proof of concept of this technique with the SPH simulation is shown in Extended Data Figure~\ref{extfig:pv-wiggle-amp:sim}\rev{b}.  
Using the analytic modeling code  \texttt{giggle}\footnote{\url{http://doi.org/10.5281/zenodo.10205110}} of Longarini et al. (2021)\cite{longarini2021} (Methods),  we calculate the minor axis PV wiggle magnitude
in gravitationally unstable disk models 
for $60\times60$ combinations of $M_{\rm disk}/M_{\star}$ and $\beta$,  letting each vary within the ranges 
$0.0 \leq M_{\rm disk}/M_{\star} \leq \frev{0.4}$ and $\frev{10^{-2}} \leq \beta \leq 10^{2}$. 
A demonstrative analytic curve for the minor axis PV \rev{w}iggle 
\rev{from} the same 
model shown in Figure~\ref{fig:2:pp-wiggle}b 
is underlaid in Figure~\ref{fig:4:amplitude}a for qualitative comparison. 
Figure~\ref{fig:4:amplitude}c shows the resulting map of $60\times60$ analytic minor axis PV wiggle magnitudes.
Overlaying contours in this map at the magnitude values measured for the AB Aur $^{13}$CO and C$^{18}$O southern minor axis PV wiggles, 
we find a disk mass in the gravitationally unstable regime with \rev{$0.1 \lesssim M_{\rm disk}/M_{\star} \lesssim 0.3$} for a cooling timescale of $0.1<\beta<10$. 
This result is robust to plausible variations in the analytic model parameter choices \rev{(Extended Data Figure Figure~\ref{extfig:effect-p-m}).}   
This disk mass range is broadly consistent with the observed spiral morphology --- a lower disk mass 
may result in a large number of more tightly wound spirals 
than we observe, and vice versa\cite{lodato-rice-2004, lodato-rice-2005}. 
\frev{To demonstrate that the implied cooling timescales are compatible with the constrained disk mass values, Figure~\ref{fig:4:amplitude}c also  
displays 
ranges of $\beta$ derived from independent radiative cooling prescriptions 
(see Methods).}

The detection of GI in the disk around AB Aur, a 
Class II YSO\cite{henning1998, bouwman2000}, demonstrates that gravitational instability can take place during later evolutionary stages. 
This result, together with previous reports of multiple protoplanet candidates \rev{in and amongst spiral arms}  
in the system\cite{oppenheimer2008-abaur, tang2017-abaur12COspirals, boccaletti2020-abaursphere, currie2022-abaurb} \rev{(Extended Data Figure \ref{extfig:candidate-sites-planets})}, 
provides a direct observational connection between gravitational instability and planet formation. 
Looking forward, the AB Aur system can be an ideal testbed for understanding how planet formation is facilitated by GI-induced spiral arms -- whether by fragmentation into gas clumps enabled by rapid cooling\cite{gammie2001, rice2003, zhu2012-challenges-clumps, deng2021-magnetic-fragmentation} ($\beta \lesssim 3$), or by dust collapse of solids concentrated within spiral arms sustained by slow cooling\cite{rice2004, longarini2023b, booth-clarke2016-dustySG, rowther2024-dustconcentration-GI} ($\beta \gtrsim 5$).



\clearpage

\clearpage

\end{bibunit}

\begin{bibunit}[sn-nature]

\clearpage
\section*{Methods} 

\noindent {\bf Additional information on the source.} 
AB Aur is accreting from the disk at a rate $\dot{M}\sim10^{-7} M_\odot$ yr$^{-1}$ (ref.\cite{salyk2013-accretion}), 
within the range 
expected for modest GI-driven accretion ($10^{-7} - 10^{-6} M_\odot$ yr$^{-1}$)\cite{rice-armitage-2009}. 
\rev{This accretion rate, taken together with the current age $t_0=2.5-4.4$ Myr\cite{vandenancker1997, dewarf2003, beck2019-H2, garufi2024-destinys-taurus}, implies a high ``latent disk mass'': $M_{\rm disk}^{\rm latent}=\dot{M}(t_0)\times t_0=0.25-0.44 M_\odot$, 
or $M_{\rm disk}^{\rm latent}/M_{\rm star}\sim0.1-0.2$. 
$M_{\rm disk}^{\rm latent}$ provides an accretion rate-based 
assessment of disk mass, assuming 
a constant stellar accretion rate $\dot{M}$ and we are observing the system mid-way through the disk's lifetime\cite{hartmann1998, dong2018-gi-or-planets}. This is a conservative estimate as the accretion rate at earlier epochs is likely higher\cite{sicilia-aguilar2010}.} 
\rev{In millimeter continuum observations, the disk shows a dust ring at $\sim1''$ and a cavity inside\cite{tang2012-abaur}, likely caused by the trapping of millimeter-sized dust at a pressure bump. The dust ring is located inside the main spirals in both the scattered light and gas emission.} 
Late infall from above or below the main disk plane\cite{nakajima-golimowski1995-palomar, grady1999-hst, tang2012-abaur, riviere2020-rosetta1} 
is likely encouraging GI by providing a source of mass to maintain a high $M_{\rm disk}/M_{\star}$ value\cite{fukagawa2004, hall2019-temporalGIspiralsALMA}.

\noindent {\bf ALMA observations.} 
We observed AB Aur with ALMA in April, May and September 2022 under ALMA program ID 2021.1.00690.S (PI: R. Dong). Measurements were taken with the Band 6 \frev{receivers\cite{ediss2004-band6}} in array configurations C-3 (2 execution blocks) and C-6 (6 execution blocks). 
In total, the 8 execution blocks reached an on-source integration time of 5.75 hours, making this the longest fine-kinematics ($v_{\rm chan}<100$ m/s) program toward a single protoplanetary disk to date. \frev{Extended Data} Table \ref{tab:observations} provides details of the observations.  
We centered one spectral window (SPW) at the $^{13}$CO $J=2-1$ molecular emission line transition rest frequency \rev{(220.3986 GHz), } 
covering a bandwidth of 58.594 MHz with 1920 channels, resulting in the highest achievable spectral resolution of 41.510 m/s after default spectral averaging with $N=2$ by Hanning smoothing within the correlator data processor. A second SPW was centered at the C$^{18}$O $J=2-1$ rest frequency \rev{(219.5603 GHz)} 
covering the same bandwidth with half as many channels (960 channels; due to sharing a baseband with another SPW), achieving a 83.336 m/s spectral resolution. 
To enable self-calibration, our correlator setup sampled the continuum in another SPW centered at $233.012$ GHz with 128 channels each 15.625 MHz in width, obtaining the full available 2.0 GHz bandwidth. 
Using the continuum data, all execution blocks were aligned to a common phase center in the \textit{uv}-plane. 
We performed a series of phase-only self-calibration iterations,  
and avoided combining by SPW in the first two rounds to remove any potential per-SPW phase offsets. 
We also carried out one round of amplitude and phase self-calibration. 
Finally, we applied the phase center realignments and calibration gain tables (that we generated with the continuum data) to the line data. We performed continuum subtraction in the \textit{uv}-plane using the \texttt{uvcontsub} task.

All imaging was performed with the {\sc CASA} \texttt{tclean} task.
We used the multiscale 
deconvolution algorithm\cite{cornwell2008} with 
(Gaussian) deconvolution scales [$0.02''$, $0.1''$, $0.3''$, $0.6''$, $1.0''$]. 
We did not image with a Keplerian mask so as not to restrict our ability to observe non-Keplerian emission. 
After experimentation with {\sc CASA}'s auto-multithresh masking algorithm\cite{kepley2020-automultithresh}, we adopted  
an imaging strategy similar to {\sc PHANGS-ALMA}\cite{leroy2021-phangsalma},  
in which we clean conservatively, with a broad mask (\texttt{usemask='pb'} and \texttt{pbmask=0.2}), forcing frequent major cycles\footnote{The $^{13}$CO robust 0.5 cube underwent 198 major cycles and the C$^{18}$O cube underwent 76.}. 
To achieve frequent major cycles we set 
the maximum number of minor cycle iterations per channel to \texttt{cycleniter=80}, 
the minor cycle threshold to \texttt{max\_psf\_sidelobe\_level=3.0} and \texttt{minpsffraction=0.5}, 
and the maximum assigned clean component to \texttt{gain=0.02} times the peak residual. 
We adopted a Briggs robust weighting scheme, and generated two sets of image cubes; one with a robust value of $0.5$ and a second with robust $1.5$. 
The corresponding beam sizes for $^{13}$CO are $237\times175$ mas, $1.2^\circ$ for robust $0.5$ and $390\times274$ mas, $-1.4^\circ$ for robust $1.5$. 
We imaged \rev{with a FOV} out to the primary beam FWHM ($38''$) with $0.02''$ pixels ($9$ or $12$ pixels per synthesized beam minor or major axis, respectively). 
We imaged in LSRK velocity channels at $42$ m/s for $^{13}$CO and $84$ m/s for C$^{18}$O respectively (nearly native channel spacing). 
The \texttt{CLEAN} threshold was set to $5\times$ the rms noise measured in $20$ line-free channels 
of the dirty image cube. 
We applied JvM correction\cite{JvM1995-correction, czekala2021-maps2} and primary beam correction. The rms \rev{noise in} the resulting $^{13}$CO \rev{cubes imaged with robust $0.5$ and robust $1.5$} is $2.0$ mJy/beam and $1.2$ mJy/beam respectively, \rev{and $0.6$ mJy/beam in the C$^{18}$O cube imaged with robust $1.5$.}

We used the robust $0.5$ image cubes for our position-position analysis (moment maps; Figures \ref{fig:1:residuals} \& \ref{fig:2:pp-wiggle}) and the robust $1.5$ cubes for our position-velocity analysis (PV diagrams and line centers; Figures \ref{fig:3:spectra} \& \ref{fig:4:amplitude}).
We made the moment 0, 1 and 2 maps 
using the \texttt{bettermoments}\cite{teague2018-bettermoments, teague2018-robust-linecentroids} methods `collapse\_zeroth, `collapse\_first', and `collapse\_percentiles', respectively. We note that we calculate our ``moment 2'' maps as the average of the red- and blue-shifted line widths about the intensity-weighted median line center (i.e., as the average of the \texttt{wpdVr} and \texttt{wpdVb} maps). Mathematically this is a different approach to find the line width than the classic moment 2 approach, though in 
our case 
we find the two yield nearly identical outcomes. 
We applied sigma-clipping at $5\times$ the rms noise and performed no spectral smoothing.

\noindent {\bf Geometric properties.} 
We used the Python package \texttt{eddy}\cite{teague2019-eddy} to infer geometric properties of the disk, namely to constrain the disk center $x_{0}$, $y_{0}$, the disk inclination $i$, the position angle ${\rm PA}$, the systemic velocity $v_{\rm sys}$, and the dynamical stellar mass $M_{\star}$. 
We performed an MCMC to fit the C$^{18}$O moment 1 map \rev{(Extended Data Figure \ref{extfig:hp-vkep-residuals})} with a geometrically thin Keplerian disk rotation profile: 
\begin{equation}
    \label{eqn:v0-eddy}
    v_{0} = \sqrt{\frac{G \, M_{\star}}{r} } \cdot \sin{i} \cdot \cos{\phi} + v_{\rm sys} \, ,
\end{equation}
where 
$r$ is the disk radius, $\phi$ is the azimuthal angle around the disk, and $G$ is the gravitational constant. 
Following convention, we fix the inclination to the value found from fitting the continuum, $i=23.2^{\circ}$ (ref.\rev{\cite{tang2012-abaur, tang2017-abaur12COspirals}}), and the distance to 155.9 pc (Gaia DR3\cite{gaia-mission-2016, gaiaDR3-2023}). 
We assumed flat priors for all values and spatially downsampled the rotation map to the beam FWHM prior to the likelihood calculation so that only spatially independent pixels were considered. 
The calculation of the posterior distributions was run with 128 walkers and an initial burn-in period of 10,000 steps before the posterior distributions were sampled for additional 10,000 steps.  
The resulting posterior distributions were 
$x_{0}= -5 \pm 7$ mas, 
$y_{0}= -17 \pm 7$ mas, 
${\rm PA} = 236.7 \pm 0.3 \, ^{\circ}$,  
$M_{\star}= 2.23 \pm 0.02 \, M_{\odot}$, and 
$v_{\rm sys} = 5858 \pm 5$ m/s, 
where we report the uncertainties represented by the 16th and 84th percentiles about the median value. 
The latter three values are consistent with constraints from previous observations\cite{beck2019-H2, tang2017-abaur12COspirals, pietu2005, dewarf2003, tang2012-abaur}.

\noindent {\bf Hydrodynamic simulations and synthetic ALMA observations.} 
We performed 3D global smoothed-particle hydrodynamic (SPH) simulations with the {\sc PHANTOM} code\cite{price2018-phantom} 
using 1 million SPH particles. 
We assumed a central star mass of $2.4 \, M_{\odot}$ (ref.\cite{dewarf2003, beck2019-H2}), 
represented by a sink particle\cite{bate1995} with accretion radius set to 60 au. The initial inner and outer disk radii were set to $r_{\rm in,SPH}=80$ au and $r_{\rm out,SPH}=500$ au, respectively. 
We set the initial gas mass to $0.7 \, M_{\odot}$, \rev{corresponding to $M_{\rm disk}/M_{\star}=0.29$.} 
\rev{The} surface density profile \rev{follows} $\Sigma \propto r^{-p}$ (where the power-law index $p=1.0$), and \rev{the} sound speed profile \rev{follows} $c_{\rm s} \propto r^{-q}$ (where $q=0.25$). The initial disk aspect ratio was set to $H/r=0.05$ at 80 au.  
We set $\alpha_{\rm SPH}$ such that $\alpha_{\rm min}\leq \alpha_{\rm SPH}  \leq \alpha_{\rm max}$, with $\alpha_{\rm min} = 0.001$ and  $\alpha_{\rm max}=1.0$, with the value of $\alpha_{\rm SPH}$ set by the Cullen \& Dehnen (2010)\cite{cullendehnen2010} 
switch that increases viscosity only in the case of converging flows. This results in a Shakura-Sunyaev viscosity of $\alpha_{\rm SS}\approx 0.01$ throughout the disk. 

We assumed an adiabatic equation of state, with heating from compressional $P \, {\rm d}V$ work and shock heating. 
The disk cools by Gammie cooling\cite{gammie2001} (a.k.a. $\beta$-cooling) where the cooling timescale is proportional to the local dynamical time by the factor $\beta$, such that $t_{\rm cool}(r) = \beta \, \Omega^{-1}(r)$, where $\Omega(r)=(G\, M_{\star} / r^{3})^{1/2}$ is the Keplerian frequency. We set $\beta=10$, a typical value used or found in simulations\cite{hall2020, terry2022-diskmass, longarini2021, panequecarreno2021-elias27}. 
We let the simulation evolve for five orbital periods of the outermost particle, at which point the disk settles into a state in which the Toomre Q parameter is between 1 and 2 from $r_{\rm in,SPH}$ to 1.1$r_{\rm out,SPH}$.

We computed the disk thermal structure and $^{13}$CO ($J=2-1$) model line cubes using the Monte Carlo radiative transfer code {\sc MCFOST}\cite{pinte2006, pinte2009}. 
We assumed the $^{13}$CO molecule is in local thermodynamic equilibrium (LTE) with its surroundings and that the dust is in thermal equilibrium with the gas ($T_{\rm gas}=T_{\rm dust}$). 
We set the $^{13}$CO/H$_{2}$ abundance to $7\times10^{-7}$ (ref.\cite{pinte2018-hd97048, hall2020}) and we used 
$\approx 10^{7}$ photon packets to calculate $T_{\rm dust}$. 
Voronoi tesselation was performed on 990,972 SPH particles which corresponded to 99\% of the mass in the simulation. 
We set the total dust mass to 1\% of the total SPH gas mass and used a dust grain population with 50 logarithmic bins ranging in size from $0.1 \, \mu$m to $3.0$ mm. 
The dust optical properties are computed using Mie theory. 
The central star was represented as a sphere of radius $2.5 \, R_{\odot}$ radiating isotropically 
at an effective temperature $T_{\rm eff}=9770$ K, 
set to match AB Aur\cite{li2016-rstar, hillenbrand1992-Teff, natta2001-Teff, currie2022-abaurb}. 
The disk was given an inclination of $23.2^{\circ}$, a position angle of $236.7^{\circ}$ (where PA is measured east of north to the red-shifted major axis), and placed at a distance of $155.9$ pc, all consistent with the AB Aur system. 

We used the same {\sc PHANTOM} simulation to create both the GI and Keplerian model line cubes shown in Figure~\ref{fig:2:pp-wiggle}c and \ref{fig:3:spectra}. We created the Keplerian counterpart with {\sc MCFOST}, using the flags \texttt{-no\_vr} and \texttt{-no\_vz} to force the radial and vertical velocities to be zero, and \texttt{-vphi\_Kep} to force the azimuthal velocities to be Keplerian. 
Both $^{13}$CO model line cubes were generated with {\sc MCFOST}, binned at the observed spectral resolution of 42 m/s,  
and gridded in the image plane to have $2048\times2048$ pixels of size $0.02''$.  
We assumed a turbulent velocity of $0.05$ km/s. 

We generated synthetic ALMA image cubes from the $^{13}$CO model line cubes using \texttt{syndisk}\footnote{\url{https://github.com/richteague/syndisk}} to match the properties of the observed AB Aur $^{13}$CO image cubes (robust $0.5$ and $1.5$). 
In the latter case the model line cube was convolved with a beam of size $0.390'' \times 0.274''$ and PA $-1.4^{\circ}$. Correlated noise was added with an rms of $1.2$ mJy/beam. 
The model data were then smoothed with a Hanning spectral response function with a resolution of 42 m/s. Effects associated with interferometric or spatial filtering are not captured by this process, and our synthetic ALMA image cubes are effectively fully-sampled in the $uv$-plane. The synthetic cubes were collapsed into moment maps following the same procedure as the AB Aur data (Extended Data Figure~\ref{extfig:momentmaps}).

\noindent {\bf Analytic modeling.} 
We analytically compute the velocity fields of gravitationally unstable disks using the \texttt{giggle}\footnote{\url{http://doi.org/10.5281/zenodo.10205110}} package developed by Longarini et al. (2021)\cite{longarini2021}. 
Working in 2D polar coordinates ($r$, $\phi$), \texttt{giggle} considers a geometrically thin disk with surface density profile $\Sigma_0 \propto r^{-p}$ and inclination $i$, centered on a star of mass $M_{\star}$. It computes the projected line-of-sight velocity field as 
\begin{equation}
    \label{eqn:projected-velocity-field}
    v_{\rm los} = (v_{r}  \sin{\phi} + v_{\phi} \cos{\phi}) \sin{i} + v_{\rm sys} \, ,
\end{equation}
where $v_{r}$ and $v_{\phi}$ are the radial and azimuthal components of the disk velocity field. 
The basic state of the disk 
(i.e., considering only the gravitational potential contribution from the central star)
is assumed to be Keplerian: $v_{r}=0$ and $v_{\phi}=v_{\rm Kep}$. 
The scheme of the model is to determine the perturbations in 
$v_{r}$ and $v_{\phi}$ generated by gravitational instability 
by taking into account the additional gravitational contribution from the disk, 
which is initialized as marginally unstable  
and 
imprinted with global spiral density perturbations.  
The model computes the velocity field under the assumption that the disc is self-regulated. 
This state is imposed by assuming a balance between heating (by compression and shocks within the spiral arms) and cooling (by radiative processes). 
As such, 
the amplitude of the spiral density perturbations $A_{\Sigma_{\rm spir}}/\Sigma_0$ 
saturated to a finite value proportional to the 
cooling timescale $\beta$:\cite{lodato2008, cossins2009}
\begin{equation}
    \label{eqn:cossins2009}
    \frac{A_{\Sigma_{\rm spir}}}{\Sigma_0}  = \chi \beta^{-1/2} \, ,
\end{equation}
where the proportionality factor $\chi$ is of order unity\cite{longarini2021, cossins2009}.
The imprinted spiral density perturbation is assumed to be small relative to the background surface density, so that all the relevant quantities (density $\Sigma$,  gravitational potential $\Phi$, velocities $v_r$ and $v_\phi$, and enthalpy $h$) can be written as a linear sum of the basic state and the perturbation: 
\begin{align}
    \label{eqn:linear-forms}
    X(r, \phi) = X_0(r) + X_{\rm spir}(r, \phi) \, . 
\end{align}
The spiral perturbation in density is given the form
\begin{equation}
    \label{eqn:spiral-density-wave}
    \Sigma_{\rm spir}(r, \phi) = \Re{ \big[ A_{\Sigma_{\rm spir}} \, e^{j (m \phi + \psi(r))}  \big] } \, ,
\end{equation} 
where $j=\sqrt{-1}$ (as we are using $i$ to represent the disk inclination), and $m$ is the azimuthal wavenumber. The ``shape function'' $\psi(r)$ is described by $m$ and the spiral pitch angle $\alpha_{\rm pitch}$ as:
\begin{equation}
    \label{eqn:shape-function}
    \psi(r)= \frac{m}{\tan{\alpha_{\rm pitch}}} \log r \, ,
\end{equation}
which is related to the radial wavenumber $k$ by $d \psi / dr = k$. The spiral density perturbation necessarily 
introduces a corresponding perturbation to the gravitational potential: 
\begin{equation}
    \label{eqn:spiral-gravitational-potential}
    \Phi_{\rm spir}(r, \phi) = -\frac{2 \pi G}{|k|} \, \Sigma_{\rm spir}(r, \phi) \, .
\end{equation}
The negative proportionality $\Phi_{\rm spir} \propto - \Sigma_{\rm spir}$ is the definition of self-gravitating spiral arms. 
As a result, corresponding perturbations in the azimuthal and radial velocities are driven: 
\begin{equation}
    \label{eqn:vr}
    v_{r}(r, \phi) = \Re{ \big[ A_{v_{r}}(r) \cdot  e^{j (m \phi + \psi(r))}  \big] } \, ,
\end{equation}
\begin{equation}
    \label{eqn:vphi}
    v_{\phi}(r, \phi) = \Re{ \big[ A_{v_{\phi}}(r) \cdot e^{j (m \phi + \psi(r))}  \big] } + r \Omega \, ,
\end{equation} 
where we note $r \Omega \neq v_{\rm Kep}$ because the angular frequency $\Omega$ includes super-Keplerian rotation from the disk mass contribution:    
\begin{equation}
    \label{eqn:gravitational-potential-field}
    \Omega^2 = \frac{G M_{\star}}{r^3} + \frac{1}{r} \frac{\partial \Phi_{\rm disk}}{\partial r} \, .
\end{equation}
By assuming the disk is marginally unstable, and by maintaining the self-regulated state condition, the amplitude of the radial and azimuthal velocity perturbations $A_{v_{r}}(r)$ and $A_{v_{\phi}}(r)$ are determined:\cite{longarini2021}  
\begin{equation}
    \label{eqn:vr-amplitude}
    A_{v_{r}}(r) = 2  j  m  \chi  \beta^{-1/2} \bigg( \frac{M_{\rm disk}(r)}{M_{\star}}\bigg)^{2} \, v_{\rm Kep}(r) \, , 
\end{equation}
\begin{equation}
    \label{eqn:vphi-amplitude}
    A_{v_{\phi}}(r) = -\frac{1}{2}  j \chi  \beta^{-1/2} \bigg( \frac{M_{\rm disk}(r)}{M_{\star}}\bigg) \, v_{\rm Kep}(r) \, , 
\end{equation}
where $M_{\rm disk}(r)$ is the disk mass enclosed within radius $r$. 
\rev{
With a surface density profile $\Sigma_0(r) \propto r^{-p}$, then $M_{\rm disk}(r) \propto r^{-p +2}$, and 
the amplitude of the radial perturbation is described by $A_{v_{r}}(r) \propto r^{-2p + 7/2}$. For $p<7/4$, $A_{v_{r}}(r)$ is an increasing function of radius. } 
\rev{The factor of imaginary number $j$ in Eqn. \ref{eqn:vr-amplitude}
has important physical consequences:    
when the real component of $A_{v_{r}}(r)$ is taken (Eqn. \ref{eqn:vr}), the radial velocity perturbation is $\pi/2$ out of phase with the spiral density perturbation (Eqn. \ref{eqn:spiral-density-wave}), and 
convergent 
at the locations where $\Sigma_{\rm spir}$ takes a maximum. 
Explicitly, 
\begin{align}
    \label{eqn:phase-offsets}
    v_r(r, \phi)\bigg\rvert_{\phi=\pi/2} \propto - \sin{\bigg(m\frac{\pi}{2} + \psi(r) } \bigg)  \, , \\
    \Sigma_{\rm spir}(r, \phi)\bigg\rvert_{\phi=\pi/2} \propto \cos{\bigg(m\frac{\pi}{2} + \psi(r) } \bigg) \, .
\end{align}}

For qualitative visual comparison with the AB Aur moment 1 map in Figure~\ref{fig:2:pp-wiggle}a, we 
compute the projected line-of-sight velocity field of a gravitationally unstable disk with $\beta=10$ and $M_{\rm disk}/M_{\star}=\rev{0.3}$  
in Figure~\ref{fig:2:pp-wiggle}b. 
We set $m=3$ and $\alpha_{\rm pitch}=15^{\circ}$ to approximately match the $^{13}$CO spirals in the AB Aur disk (Figure~\ref{fig:1:residuals}), and assume 
$p=1.0$ and $\chi=1.0$ (ref.\cite{cossins2009}). 
The dominant azimuthal wavenumber is expected to be inversely related to the disk-to-star mass ratio $q$, roughly obeying $m \sim 1/q$ (ref.\cite{lodato-rice-2004, cossins2009, dong2015-GIspirals-scatteredlight}), so our choice of $m=3$ is consistent with $M_{\rm disk}/M_{\star}\approx 0.3$.

\noindent {\bf Revealing global spiral structure.} 
We obtain the residual moment maps shown in Figure~\ref{fig:1:residuals} using a variation on the conventional high-pass filtering (a.k.a. unsharp masking) technique. The conventional method is to convolve the image with a Gaussian kernel and subtract the blurred image from the original. It is a common technique to \rev{increase the visual contrast of variations in an image and has been used successfully to} reveal spiral structure disks (e.g.\cite{boccaletti2020-abaursphere, rosotti2020-hd100453, perez2016-elias27, meru2017-elias27, zhang2023-destinys, norfolk2022-hd100546, garufi2024-destinys-taurus}). Here, 
we perform the convolution with a radially expanding kernel\footnote{\url{https://github.com/jjspeedie/expanding\_kernel}} -- that is, with a Gaussian kernel whose FWHM, $w$, increases with radial distance from the image center (i.e., with disk radius) with a simple power-law dependence: 
\begin{equation}
    \label{eqn:expanding_kernel}
    w(r) = w_{0} \cdot (r/r_0)^{\gamma} \, ,
\end{equation}
where $w_{0}$ is the kernel width at $r_0=1''$. 
\rev{A radially expanding kernel provides a way to highlight variations more evenly throughout the disk, given the spatial scales of the variations --which are expected to track with the local scale height and increase with radius-- and the dynamical range of the variations, which fall with radius. After experimentation we adopt $w_{0}=0.3''$ and $\gamma=0.25$, though we emphasize this is a qualitative choice and 
the key spiral features, such as their locations, are robust against a variety of choices in kernel parameters.} 
\rev{The high-pass filter technique is also flexible to the disk emission surface morphology, and can capture global scale deviations from Keplerian rotation in the background disk.}  Extended Data Figure~\ref{extfig:hp-vkep-residuals} compares the residual \rev{moment 1 maps in $^{13}$CO and C$^{18}$O} obtained after subtracting \rev{the} axisymmetric geometrically thin Keplerian model (Eqn. \ref{eqn:v0-eddy}) vs. after subtracting a blurred version of the moment 1 map made with the expanding kernel filter. 
The Keplerian residuals (\rev{panels c and h}) show signs of global scale deviation from Keplerian: the east (west) side is generally blue-shifted (red-shifted), hinting at 
super-Keplerian rotation, signatures of disk mass contributing to the total mass of the system. 
While spiral structure is indeed also visible in the Keplerian residuals, 
the expanding kernel residuals (\rev{panels e and j}) reveal the underlying spiral structure in a spatially even manner, indicating that the \rev{expanding kernel} background model (\rev{panels d and i}) \rev{more} successfully captures the \rev{quasi-local background disk velocity.} 
We note that this background model is non-axisymmetric; 
\rev{it displays excess blueshifted velocity in the southeast quadrant of the disk such that} 
the contour of $v_{\rm los}=v_{\rm sys}$ diverges \rev{westward} from the minor axis south of the star, possibly indicative of a global disk warp. This 
is what necessitates a detrending of the line centers to isolate the sinusoidal component of the southern minor axis PV wiggle in Figure~\ref{fig:4:amplitude}a (see section ``Measuring the magnitude of AB Aur’s minor axis PV wiggle''). 
Filtered moment maps for the synthetic ALMA observations of the simulated SPH GI disk are shown in Extended Data Figure~\ref{extfig:momentmaps-filtered}.

\noindent {\bf Global kinematics of self-gravitating spiral arms.}  
Radially convergent motion (as in Figure \ref{fig:1:residuals}bcd insets) serves as a kinematic signature for the location of self-gravitating spiral arms at disk azimuths where the radial velocity perturbation contributes sufficiently strongly to the observed velocity field, and thus cannot be a fully unambiguous locator at disk azimuths away from the minor axis. 
Extended Data Figure \ref{extfig:kinematics-spirals}c and g 
provide maps of velocity residuals from Keplerian for the 2D analytic GI disk model and the SPH GI disk simulation. 
The convergent motion toward the spiral spines is visible for a range of azimuths around the minor axis, but becomes progressively less clear moving toward the major axis as the azimuthal velocity --super-Keplerian rotation-- contributes progressively more to the line-of-sight. 
However, high-pass filtering (panel h) captures and removes the background super-Keplerian rotation, leaving a residual map that resembles the isolated radial component (panel d). 
Extended Data Figure \ref{extfig:kinematics-spirals}i-l overlays the locations of $^{13}$CO spirals in the AB Aur disk (from filtered moment 0/2; Figure \ref{fig:1:residuals}cd) onto the filtered moment 1 maps, in order to illustrate where convergent motion does or does not serve as a locator throughout the disk. 
Ambiguity occurs around the major axis, which is a location of transition in the sign of $v_{r} \sin{i} \sin{\phi} $ (first term of Eqn. \ref{eqn:projected-velocity-field}), and when two spirals are not well separated and their motions superimpose.
Three of the seven spiral structures in VLT/SPHERE scattered light appear to be spatially associable 
with those in $^{13}$CO (S1, S5, S7; panel l inset). 
Offsets in the southeast quadrant of the disk (S2, S3, S4) may be further indication of a disk warp (Extended Data Figure \ref{extfig:hp-vkep-residuals}di), or other non-trivial phenomena (e.g., vertical density and temperature gradients, projection effects\cite{ginski2016-scatteredlight}).

The kinematic signatures observed in the present ALMA dataset --probing disk scales $\sim$100 to 1,000 au-- are recognizably different from what is expected for planet-driven perturbations. 
Planetary wakes are dampened and become nearly circular as they propagate away from the planet\cite{goodman-rafikov2001, rafikov2002, ogilvie-lubow2002}, whereas GI-driven spirals maintain their modest pitch angles with radius and the amplitude of the induced velocity perturbations depends on the enclosed disk mass (Eqns. \ref{eqn:vr-amplitude} \& \ref{eqn:vphi-amplitude}). 
In the planetary case, the density and radial velocity perturbations are in phase (their peaks spatially coincide), 
and the pattern of motion within an arm along a radial cross-section is divergent\cite{bollati2021-theory-kinks, hilder2023-wakeflow-joss}.  
Overall, the essential characteristic of GI-induced spirals is that they occur globally\cite{hall2020, longarini2021} (c.f. Figure \ref{fig:1:residuals}, Extended Data Figures \ref{extfig:momentmaps-filtered}, \ref{extfig:global-wiggles} \& \ref{extfig:kinematics-spirals}). 
In previous datasets probing smaller spatial scales --within the AB Aur disk's central cavity-- 
planetary candidates P1/f1 (ref.\cite{tang2017-abaur12COspirals, boccaletti2020-abaursphere}), P2/b (ref.\cite{tang2017-abaur12COspirals, currie2022-abaurb, zhou2023-abaurb, biddle2024-pabeta-ABAur, currie2024-pabeta-ABAur}), and f2 (ref.\cite{boccaletti2020-abaursphere}) are known to be associated with 
--or driving-- 
spiral arms, as observed in VLT/SPHERE scattered light and/or ALMA $^{12}$CO emission. 
As shown in Extended Data Figure \ref{extfig:candidate-sites-planets}, 
due to their small separations ($\lesssim 0.7''$), 
kinematic signatures from these candidates are 
inaccessible to our ALMA observations. 
Clump-like signals `c' and `d'  
seen by HST/STIS (ref.\cite{currie2022-abaurb}) 
at wide separations ($\sim 2.75''$ and $\sim 3.72''$ respectively)
are in locations 
tentatively suggestive of constituting spiral arm fragments 
and may warrant further investigation.

\noindent {\bf Position-velocity analysis.} 
We use the robust $1.5$ image cubes for our position velocity analysis to maximize the recovery of emission at large disk radii.  
\rev{Owing to the clear association with a self-gravitating spiral arm (Figure \ref{fig:1:residuals}bcd insets), we target the wiggle on the southern minor axis. A clear spiral arm in moment 0/2 crossing the northern minor axis is also observed, but at the outer edge of the recovered $^{13}$CO and C$^{18}$O emission 
($\sim 3''$; c.f. Extended Data Figure \ref{extfig:kinematics-spirals}kl).} 
We obtain the position-velocity diagrams 
shown in Figure~\ref{fig:3:spectra} using 
\texttt{eddy}\cite{teague2019-eddy} 
to extract spectra from pixels within 
a $0.5^{\circ}$-wide 
\rev{wedge-shaped mask} 
oriented $90^{\circ}$ clockwise of the red-shifted major axis \rev{(shown in Figure~\ref{fig:3:spectra} insets)}. 
Our quantitative analysis of the minor axis PV wiggles is performed with maps of the line centers made  
using the quadratic method of 
\texttt{bettermoments}\cite{teague2018-bettermoments, teague2018-robust-linecentroids}, 
which fits a quadratic curve to the spectrum in each pixel of the cube: 
\begin{equation}
    \label{eqn:line-center-quadratic}
    I(v) = a_0 + a_1 (v-v_{\rm peak}) + a_2 (v - v_{\rm peak})^2 \, ,
\end{equation}
where $v_{\rm peak}$ is the channel of peak intensity in the spectrum. 
We select this approach over the traditional intensity-weighted mean velocity (moment 1) method specifically for its ability to provide well characterized, statistically meaningful uncertainties on the line center,  $\sigma_{v \rm los}$ (ref.\cite{teague2018-robust-linecentroids}). 
The statistical uncertainty on each line center is computed as:
\begin{equation}
    \label{eqn:line-center-sigma}
    \sigma_{v \rm los} = \sqrt{\frac{\sigma_{I}^2}{8} \bigg(\frac{3}{{a_2}^2} + \frac{{a_1}^2}{{a_2}^4} \bigg)} \, ,
\end{equation}
where $\sigma_{I}$ is the rms noise of the intensities (see ref.\cite{teague2018-robust-linecentroids} for a derivation). 
The quadratic method also has the advantage of being unaffected by sigma-clipping and of automatically distinguishing the front side of the disk from the back side\cite{teague2018-robust-linecentroids}. 
Prior to the quadratic fitting we spectrally smooth the data with a Savitzky-Golay filter of polynomial order $1$ and filter window length of $10$ channels ($420$ m/s) in the case of $^{13}$CO and $3$ channels ($252$ m/s) in the case of C$^{18}$O. The former was also applied to the two synthetic ALMA $^{13}$CO image cubes generated from the SPH simulations.  
We 
extract the values from the resulting line center and line uncertainty maps within the same wedge mask described above. The extracted line center values are shown as yellow points in Figure~\ref{fig:3:spectra} and the uncertainties are shown as yellow shaded regions in Figure~\ref{fig:4:amplitude}a.

\noindent {\bf Measuring the magnitude of \rev{the} minor axis PV wiggle.}   
Following Longarini et al. (2021)\cite{longarini2021}, we measure the `magnitude' of \rev{a} minor axis PV wiggle as the standard deviation of the line center values over a radial range. 
\rev{Bounded by the inner central cavity and the outer edge of C${^{18}}$O emission, we adopt a radial range of $1.0''$ to $5.0''$.} 
We estimate the uncertainty on \rev{the magnitude measurement} using a resampling procedure: we take 10,000 draws from Gaussian distributions centered on the observed line centers with standard deviation $\sigma_{v \rm los}$ (Eqn. \ref{eqn:line-center-sigma}) to create 10,000 instances of the minor axis PV wiggle; we compute their magnitudes; and then report the \rev{uncertainty as the} standard deviation of those 10,000 magnitude estimates. 

\rev{In addition to the wiggle, the $^{13}$CO and C$^{18}$O emission on the southern disk minor axis also exhibit an underlying monotonic blueward trend with disk radius, seen in Figure~\ref{fig:3:spectra}ab as a subtle downward bend with radius of the line centers, or equivalently in Figure~\ref{fig:2:pp-wiggle}a as a westward or clockwise shift in the contour of $v_{\rm los}=v_{\rm sys}$. We earmark this feature as a possible disk warp (Extended Data Figure~\ref{extfig:hp-vkep-residuals}di), and adopt a least-squares fitting approach to isolate the sinusoidal component of the PV wiggle. 
This approach yields the background trendline that minimizes the standard deviation of the residuals, thus providing the most conservative estimate for the magnitude of the detrended PV wiggle. 
We fit a quadratic trendline (Extended Data Figure~\ref{extfig:comment-2D-convergence}a) as it more closely resembles the high-pass filter background curve than a linear one (Extended Data Figure~\ref{extfig:comment-2D-convergence}bc). 
We show the quadratically-detrended PV wiggles in Figure~\ref{fig:4:amplitude}a and report their magnitudes in Figure~\ref{fig:4:amplitude}b. 
} 
\rev{We find very similar magnitudes for both the $^{13}$CO and C$^{18}$O wiggles, despite C$^{18}$O likely tracing lower optical depths in the AB Aur disk. This empirically 
substantiates comparisons with the 2D analytic model (next section).} 

Performing the same procedure \rev{outlined above} on the synthetic $^{13}$CO minor axis PV wiggle of the GI disk in the SPH simulation,  
we find a wiggle magnitude of $\rev{39.1} \pm 1.9$ m/s   
(Extended Data Figure~\ref{extfig:pv-wiggle-amp:sim}).

\noindent{\bf Constraining disk mass with quantitative comparisons to analytic models.}
We perform quantitative comparisons between the observed $^{13}$CO and C$^{18}$O minor axis PV wiggles 
and the projected radial velocity component in our analytic model, $v_r 
\sin{i}$ (ref.\cite{longarini2021}). From Eqns. \ref{eqn:vr} and \ref{eqn:vr-amplitude}, the projected radial velocity on the minor axis ($\phi=\pi/2$) is: 
\begin{equation}
    \label{eqn:analytic-PV-wiggle}
    v_r(r, \phi)\bigg\rvert_{\phi=\pi/2} \cdot \sin{i} = -2 m \chi \beta^{-1/2} \bigg(\frac{M_{\rm disk}(r)}{M_{\star}} \bigg)^2 \,  v_{\rm Kep}(r) \, \sin{\bigg(m\frac{\pi}{2} + \psi(r) } \bigg)\, \cdot \sin{i} \, .
\end{equation} 
\rev{
This curve reflects the disk mass enclosed within the inner and outer radii of the model, which we set to span the same projected radial range as the observed PV wiggles ($1''$ to $5''$).} 
We compute $3600$ of these curves for a $60\times60$ grid of \rev{models with (total enclosed)} $M_{\rm disk}/M_{\star}$ linearly spaced $\in [0.0, \frev{0.4}]$ and $\beta$ logarithmically spaced $\in [\frev{10^{-2}}, 10^{2}]$. 
Again we set $m = 3$ and $\alpha_{\rm pitch} = 15^{\circ}$ to match the AB Aur disk, 
and assume $p = 1.0$ and $\chi = 1.0$ (ref.\cite{cossins2009}). 
For qualitative comparison, we plot an example analytic minor axis PV wiggle behind the data in Figure~\ref{fig:4:amplitude}a; the model has $\beta=10$ and $M_{\rm disk}/M_{\star}=\rev{0.3}$. 
We show in \rev{Extended Data Figure Figure~\ref{extfig:effect-p-m}}   that $m=3$ reproduces the observed wiggle\rev{s} better than other choices, and that $p=1.5$ could also provide a satisfying match, while $p=2.0$ is too steep. \rev{Since the wiggle amplitude is independent of $\alpha_{\rm pitch}$ (Eqn. \ref{eqn:vr-amplitude}), the magnitude is constant with $\alpha_{\rm pitch}$ when sampled over the same range in phase (not shown).}  

We measure the minor axis PV wiggle magnitude of the $3600$ models 
\rev{and} 
present the resulting magnitude map in Figure~\ref{fig:4:amplitude}c. 
By drawing contours in the Figure~\ref{fig:4:amplitude}c map at the magnitude values measured for AB Aur \rev{($37.4 \pm 2.9$ m/s in $^{13}$CO and $44.2 \pm 1.3$ m/s in C$^{18}$O)}, 
we find every combination of $M_{\rm disk}/M_{\star}$ and $\beta$ that satisfy the observations. 
Repeating this procedure with our synthetic ALMA observations of the SPH GI disk simulation shown in Figure~\ref{fig:3:spectra}\rev{c}, we find that 
this technique 
successfully recovers the disk mass set in the underlying SPH simulation (Extended Data Figure~\ref{extfig:pv-wiggle-amp:sim}).

\frev{For independent physical estimates of 
plausible $\beta$ values between $1''$ to $5''$ ($155$ to $780$ au), we rely on 
radiative cooling prescriptions\cite{zhu2015-cooling, zhang-zhu2020-SG-beta}.  
From Equation 39 of Zhang \& Zhu (2020)\cite{zhang-zhu2020-SG-beta}, $\beta$ is a function of $r$ and depends on $M_{\rm disk}$ through the surface density $\Sigma$. 
We assume $T = (\frac{\phi \, L_{\star}}{8 \pi \, r^2 \, \sigma_{SB}})^{1/4}$, where $\sigma_{SB}$ is the Stefan-Boltzmann constant, $L_{\star} = 59 \, L_{\odot}$ is the stellar luminosity of AB Aur\cite{currie2022-abaurb}, and $\phi=0.02$ represents the flaring angle\cite{dullemond2018-dsharp6}.   
We use the DSHARP Rosseland mean opacity\cite{birnstiel2018-dsharp-opac} 
$\kappa_{\rm R}=\kappa_{\rm R}(T, a_{\rm max})$ for a power-law grain size distribution truncated at $a_{\rm max}$.  
We set $a_{\rm max}$ to $0.1$ mm and
the dust-to-gas mass ratio to $f=0.1\%$, 
based on radial drift arguments and lack of (sub-)mm emission at these large radii. 
We compute a $\beta(r)$ profile for each $M_{\rm disk}/M_{\star} \in [0.0, \frev{0.4}]$ and extract the values at $1''$ and $5''$. We overlay the resulting $\beta(M_{\rm disk}/M_{\star})$ ranges as white shaded regions in Extended Data Figure~\ref{extfig:effect-p-m} (where the dependence on $p$ arises from the dependence on $\Sigma$), and in Figure~\ref{fig:4:amplitude} as white horizontal bars at a selection of $M_{\rm disk}/M_{\star}$ values. 
For example, for $M_{\rm disk}/M_{\star}=0.2$ and $p=1.0$, we find $\beta(1'')=5.3$ and $\beta(5'')=3.6\times 10^{-2}$.  
While knowledge of cooling in disks is very limited, these estimates help to emphasize that not all values of $\beta$ are equally likely. 
}

\subsection{Data availability} 
All observational data products presented in this work are available through the \href{https://www.canfar.net/en/docs/digital_object_identifiers/}{CANFAR Data Publication Service} at \url{https://doi.org/10.11570/24.0087}. This includes final reduced and calibrated ALMA measurement sets, image cubes and moment maps, and processed SPHERE data. 
\frev{All simulated data products including hydrodynamic simulations and synthetic ALMA data are available at \url{https://doi.org/10.5281/zenodo.11668694}.} 
The raw ALMA data are publicly available via the ALMA archive \url{https://almascience.nrao.edu/aq/} under project ID 
2021.1.00690.S. 
The raw VLT/SPHERE data are publicly available via the ESO Science Archive Facility \url{https://archive.eso.org/eso/eso_archive_main.html} under programme 0104.C-0157(B). 

\subsection{Code availability} 
ALMA data reduction and imaging scripts are available at \url{https://jjspeedie.github.io/guide.2021.1.00690.S}. 
The Python packages used in this work are available: 
\texttt{bettermoments} (\url{https://github.com/richteague/bettermoments}), 
\texttt{eddy} (\url{https://github.com/richteague/eddy}), 
\texttt{giggle} v0 (\url{http://doi.org/10.5281/zenodo.10205110}),  
{\sc PHANTOM} (\url{https://github.com/danieljprice/phantom}), 
{\sc MCFOST} (\url{https://github.com/cpinte/mcfost}).


\clearpage




\begin{addendum}

  \item \rev{We thank our referees for their careful and insightful comments that improved the manuscript.} 
  \frev{We thank Kaitlin Kratter for enlightening discussions and valuable suggestions.} 
  J.S. thanks Ryan Loomis, Sarah Wood and Tristan Ashton at the North American ALMA Science Center (NAASC) for providing science support and technical guidance on the ALMA data 
  as part of a Data Reduction Visit to the NAASC, which was funded by the NAASC. 
  The reduction and imaging of the ALMA data was performed on NAASC computing facilities. 
  J.S. thanks Christophe Pinte, Daniel Price and Josh Calcino for support with {\sc MCFOST}, Luke Keyte and Francesco Zagaria for discussions on self-calibrating ALMA data, and Chris White for sharing perceptually uniform colormaps. 
  J.S. acknowledges financial support from the Natural Sciences and Engineering Research Council of Canada (NSERC) through the Canada Graduate Scholarships Doctoral (CGS D) program. 
  R.D. acknowledges financial support provided by the Natural Sciences and Engineering Research Council of Canada through a Discovery Grant, as well as the Alfred P. Sloan Foundation through a Sloan Research Fellowship. 
  C.L. and G.L. acknowledge funding from the European Union’s Horizon 2020 research and innovation programme under the Marie Sklodowska-Curie grant agreement \# 823823 (RISE DUSTBUSTERS project). C.L. acknowledges funding from UK Science and Technology research Council (STFC) via the consolidated grant ST/W000997/1.
  B.V. acknowledges funding from the ERC CoG project PODCAST No 864965. 
  Y.W.T. acknowledges support through NSTC grant 111-2112-M-001-064- and 112-2112-M-001-066-.
  J.H. was supported by JSPS KAKENHI Grant Numbers 21H00059, 22H01274, 23K03463. 
  This paper makes use of the following ALMA data: ADS/JAO.ALMA\#2021.1.00690.S. ALMA is a partnership of ESO (representing its member states), NSF (USA) and NINS (Japan), together with NRC (Canada), MOST and ASIAA (Taiwan), and KASI (Republic of Korea), in cooperation with the Republic of Chile. The Joint ALMA Observatory is operated by ESO, AUI/NRAO and NAOJ. The National Radio Astronomy Observatory is a facility of the National Science Foundation operated under cooperative agreement by Associated Universities, Inc.
  This work has made use of data from the European Space Agency (ESA) mission {\it Gaia} (\url{https://www.cosmos.esa.int/gaia}), processed by the {\it Gaia} Data Processing and Analysis Consortium (DPAC, \url{https://www.cosmos.esa.int/web/gaia/dpac/consortium}). Funding for the DPAC has been provided by national institutions, in particular the institutions participating in the {\it Gaia} Multilateral Agreement. 
  Based on data products created from observations collected at the European Organisation for Astronomical Research in the Southern Hemisphere under ESO programme 0104.C-0157(B). 
  This work has made use of the SPHERE Data Centre, jointly operated by OSUG/IPAG (Grenoble), PYTHEAS/LAM/CESAM (Marseille), OCA/Lagrange (Nice), Observatoire de Paris/LESIA (Paris), and Observatoire de Lyon.
  \frev{This research used the Canadian Advanced Network For Astronomy Research (CANFAR) operated in partnership by the Canadian Astronomy Data Centre and The Digital Research Alliance of Canada with support from the National Research Council of Canada the Canadian Space Agency, CANARIE and the Canadian Foundation for Innovation.}
 
  \item[Author Contributions] 
  R.D. led the ALMA proposal. 
  J.S. processed the ALMA data. 
  J.H. processed the VLT/SPHERE data. 
  C.H. performed the SPH simulations.
  J.S. performed the radiative transfer calculations. 
  C.L. and G.L. developed the analytic model. 
  J.S. performed all presented analyses. 
  J.S. and R.D. wrote the manuscript.
  All co-authors provided input to the ALMA proposal and/or the manuscript.

  \item[Competing Interests] The authors declare that they have no competing financial interests.
  
  \item[Correspondence] Correspondence and requests for materials should be addressed to: 

    J.S.~(email: jspeedie@uvic.ca),     
    R.D.~(email: rbdong@uvic.ca). 
 
\end{addendum}

\clearpage


\section*{Extended Data} 
\renewcommand{\figurename}{\textbf{Extended Data Figure}}
\renewcommand{\tablename}{\textbf{Extended Data Table}}
\setcounter{figure}{0}

\begin{table}
  \caption{ {\footnotesize Details of the ALMA Band 6 observations.} }
  \vspace{0.5em}
  \label{tab:observations}
  {\footnotesize
  \begin{tabular}{cccccccc}
  \hline
    \textbf{UTC Date}   & \textbf{Time on source $^{a}$} & \textbf{$N_{\rm ant}$} & \textbf{Baselines} & \textbf{pwv}  & \multicolumn{3}{c}{\textbf{Calibrators}}      \\
               & (min)          &      & (m)       & (mm) & Bandpass   & Flux       & Phase      \\
   \hline
    2022-04-19 & 41             & 41   & 14-500    & 0.74 & J0510+1800 & J0510+1800 & J0438+3004 \\
    2022-05-15 & 32.4           & 45   & 15-680    & 0.69 & J0510+1800 & J0510+1800 & J0438+3004 \\
   \hline
    2022-07-17 $^{b}$ & 47.4           & 41   & 15-2617   & 0.24 & J0510+1800 & J0510+1800 & J0438+3004 \\
    2022-07-19 & 47.4           & 42   & 15-2617   & 1.49 & J0510+1800 & J0510+1800 & J0438+3004 \\
    2022-07-20 & 34.8           & 44   & 15-2617   & 2.85 & J0510+1800 & J0510+1800 & J0438+3004 \\
    2022-07-21 & 47.4           & 42   & 15-2617   & 2.37 & J0510+1800 & J0510+1800 & J0438+3004 \\
    2022-07-22 & 47.3           & 44   & 15-2617   & 0.84 & J0510+1800 & J0510+1800 & J0438+3004 \\
    2022-07-22 & 47.4           & 41   & 15-2617   & 0.63 & J0510+1800 & J0510+1800 & J0438+3004 \\
  \hline
  \end{tabular}
  \par
  \begin{itemize}
    \item[$^{a}$] Total time on source: 5.75 hours (4.53 hours in C-6 and 1.22 hours in C-3).
    \item[$^{b}$] This execution block was used as the reference for phase alignment during post-processing.
  \end{itemize}
  }
\end{table}

\begin{figure}[ht]
\includegraphics[width=\textwidth,angle=0]{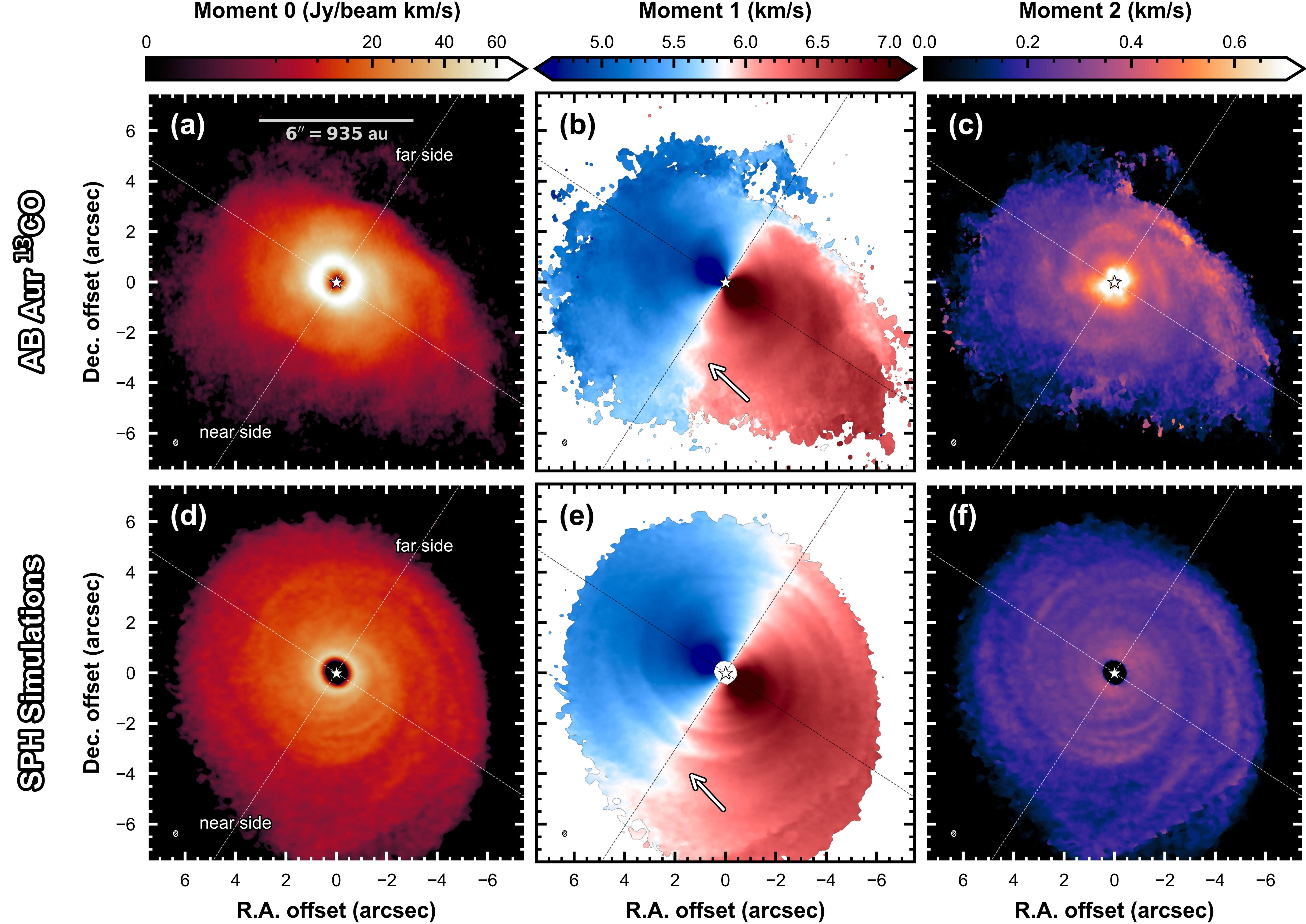}
\caption{{\bf Moment maps: AB Aur observations and GI disk simulations.} 
\textbf{(a-c)} Integrated intensity (moment 0), intensity-weighted mean velocity (moment 1), and intensity-weighted line width (moment 2) maps for the ALMA $^{13}$CO observations toward AB Aur. 
Panel (b) appears in the main text as Figure~\ref{fig:2:pp-wiggle}a. 
\textbf{(d-f)} Moment 0, 1, and 2 maps for the synthetic ALMA $^{13}$CO observations of the SPH GI disk simulation. 
Like the AB Aur observations, the simulated GI disk displays a prominent GI wiggle along the southern minor axis (indicated by white arrows). 
}
\label{extfig:momentmaps}
\end{figure}

\begin{figure}[ht]
\includegraphics[width=\textwidth,angle=0]{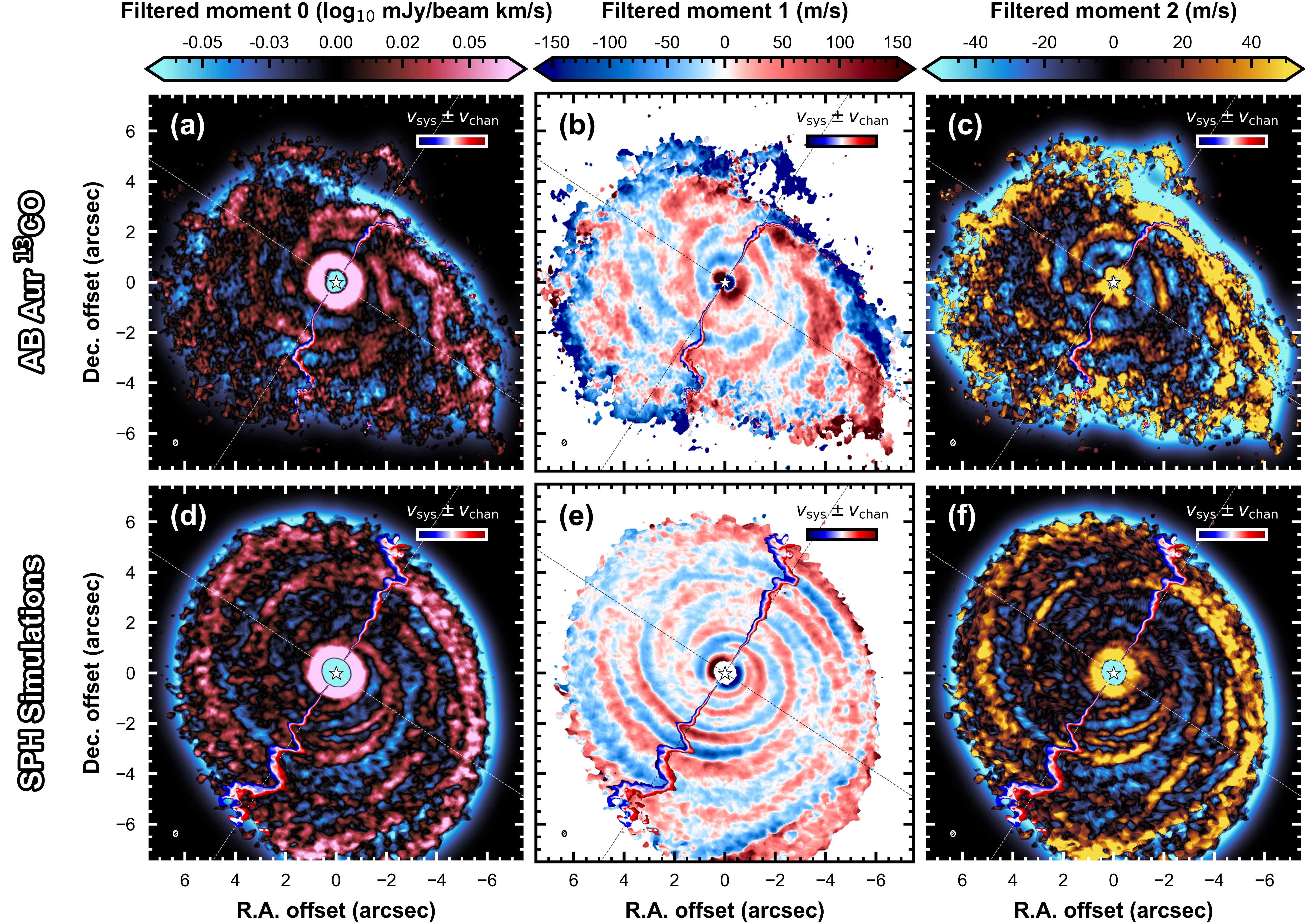}
\caption{{\bf Filtered moment maps: AB Aur observations and GI disk simulations.} Expanding kernel filter residuals of the maps shown in Extended Data Figure~\ref{extfig:momentmaps}, highlighting global spirals and velocity disturbances generated by GI.  
Panels (a-c) appear in the main text as Figure~\ref{fig:1:residuals}b-d. 
The minor axis GI wiggle indicated by white arrows in Extended Data Figure~\ref{extfig:momentmaps}b and e is shown here as \rev{an isovelocity} contour at $v_{\rm los} = v_{\rm sys} \pm v_{\rm  chan}$ in all panels. 
}
\label{extfig:momentmaps-filtered}
\end{figure}

\begin{figure}[ht]
\includegraphics[width=\textwidth,angle=0]{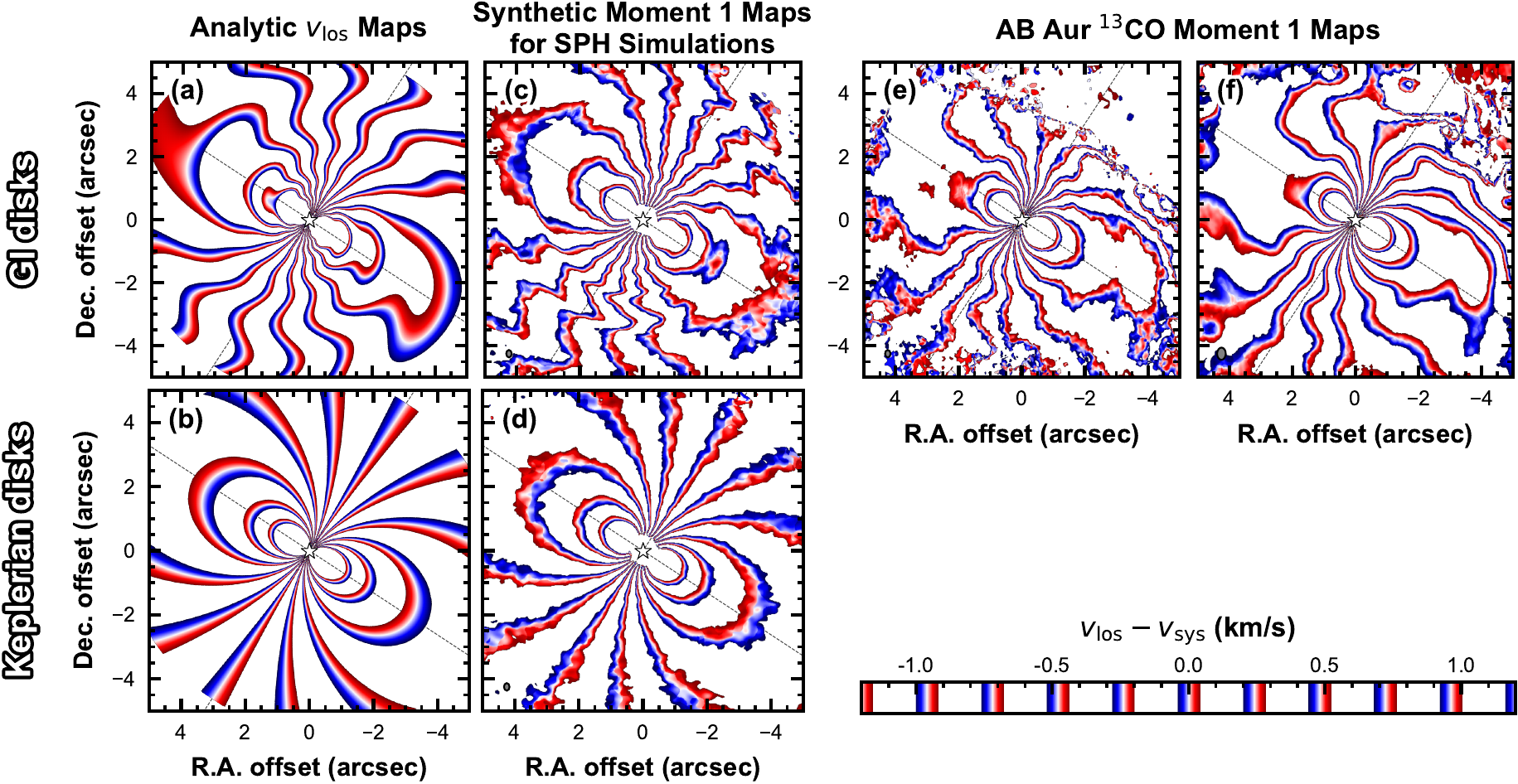}
\caption{\textbf{Global GI wiggles in analytic models, SPH simulations, and the AB Aur disk.} 
\rev{Isovelocity contours in line-of-sight velocity maps at  
the velocity values indicated by the colour bar. 
\textbf{(a)} $v_{\rm los}$ map of the 2D analytic GI disk model (shown in Figure~\ref{fig:2:pp-wiggle}b). 
\textbf{(b)} $v_{\rm los}$ map of the 2D analytic Keplerian disk model (shown in Figure~\ref{fig:2:pp-wiggle}b inset). 
\textbf{(c)} Synthetic ALMA $^{13}$CO moment 1 map for the 3D SPH GI disk simulation (shown in Figure~\ref{fig:2:pp-wiggle}c). 
\textbf{(d)} Synthetic ALMA $^{13}$CO moment 1 map for the 3D SPH Keplerian disk simulation (shown in Figure~\ref{fig:2:pp-wiggle}c inset). 
\textbf{(e)} Observed ALMA $^{13}$CO moment 1 map for the AB Aur disk, imaged with robust $0.5$ (shown in Figure~\ref{fig:2:pp-wiggle}a). 
\textbf{(f)} Like (e), but imaged with robust $1.5$. 
}
}
\label{extfig:global-wiggles}
\end{figure}

\begin{figure}[ht]
\includegraphics[width=\textwidth,angle=0]{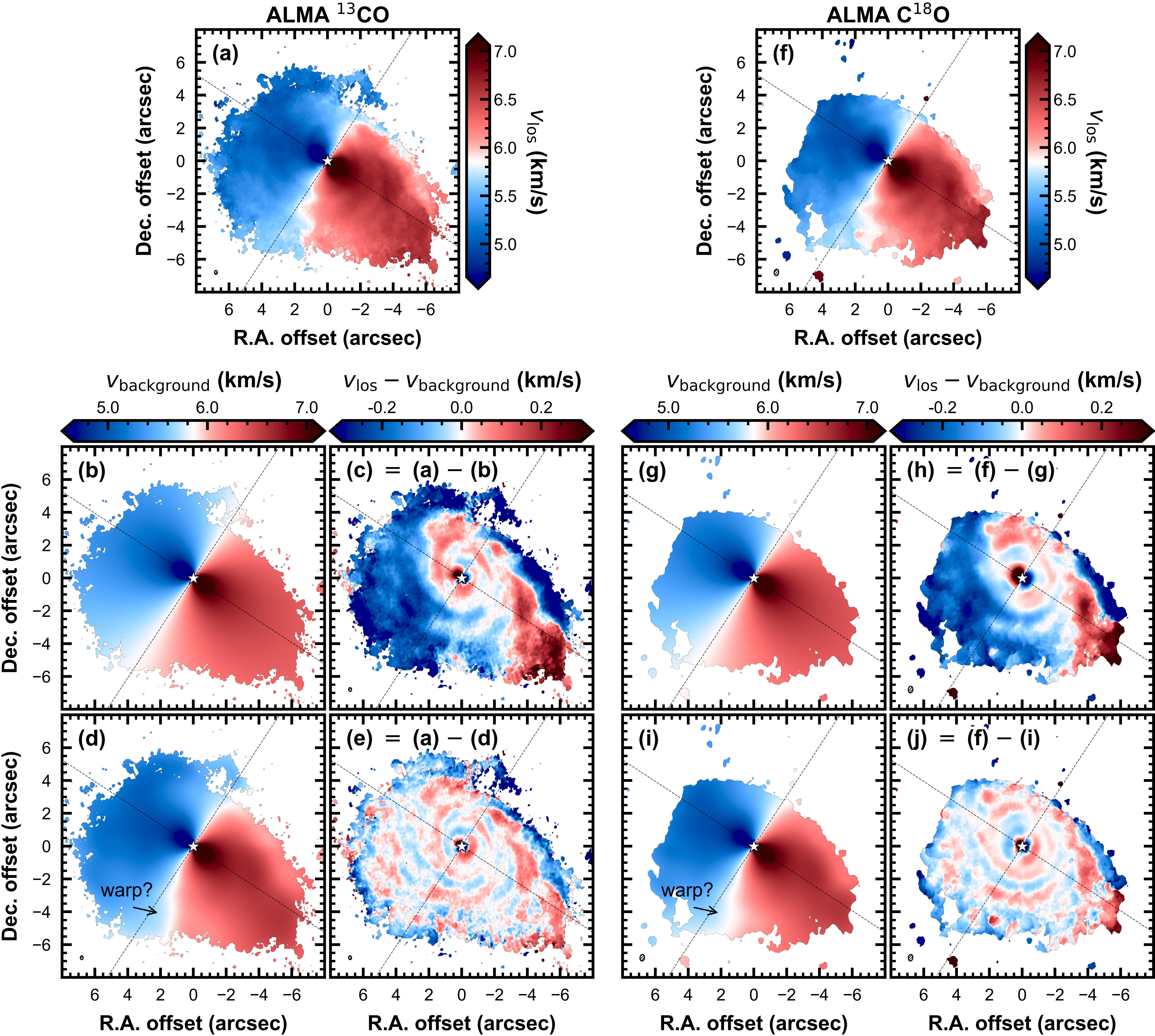}
\caption{{\bf Obtaining velocity residuals in the AB Aur disk.} 
\textbf{(a)} ALMA $^{13}$CO moment 1 map, \rev{imaged with robust $0.5$}, as shown in Figure~\ref{fig:2:pp-wiggle}a. 
\textbf{(b)} Background model made with a Keplerian rotation profile, assuming a geometrically thin axisymmetric disk \rev{(Eqn. \ref{eqn:v0-eddy})}. 
\textbf{(c)} Velocity residuals after subtracting the model in panel (b). Global spiral substructure is visible, but unevenly so. The model does not capture the non-axisymmetric emission surface morphology and/or super-Keplerian rotation. 
\textbf{(d)} Background model made with the expanding kernel filter \rev{(Eqn. \ref{eqn:expanding_kernel})}.   
\textbf{(e)} Velocity residuals after subtracting the model in panel (d), as shown in Figure~\ref{fig:1:residuals}b. 
\rev{
\textbf{(f-j)} Like (a-e) but with the ALMA C$^{18}$O moment 1 map, imaged with robust $1.5$. 
} 
}
\label{extfig:hp-vkep-residuals}
\end{figure}

\begin{figure}[ht]
\centering
\includegraphics[width=0.9\textwidth,angle=0]{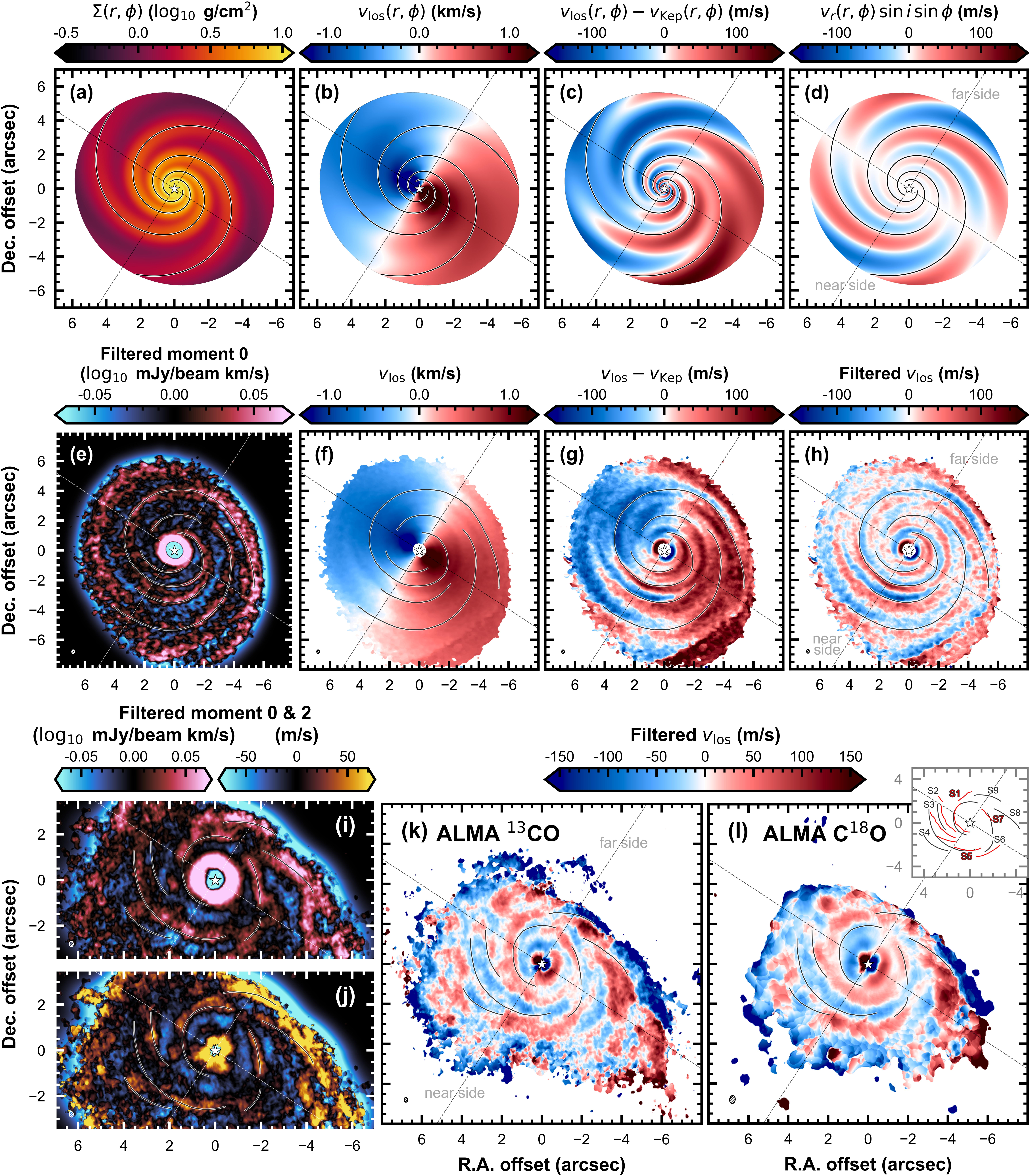}
\caption{\rev{{\bf Kinematics of GI-driven spiral arms.} 
\textbf{(a-d)} 2D analytic modeling (Longarini et al. 2021)\cite{longarini2021}. \textbf{(e-h)} Synthetic ALMA $^{13}$CO observations of the 3D SPH GI disk simulation. 
\textbf{(i-l)} ALMA observations of the AB Aur disk. 
\textbf{(a)} Disk surface density (Eqn. \ref{eqn:linear-forms} \& \ref{eqn:spiral-density-wave}). 
\textbf{(b)} Line-of-sight velocity (Eqn. \ref{eqn:projected-velocity-field}), as in Figure~\ref{fig:2:pp-wiggle}b. 
\textbf{(c)} Velocity residuals from Keplerian (i.e., subtracting Figure~\ref{fig:2:pp-wiggle}b inset). 
\textbf{(d)} Line-of-sight component of the radial velocity (first term of Eqn. \ref{eqn:projected-velocity-field}). 
\textbf{(e)} Filtered moment 0. 
\textbf{(f)} Moment 1. 
\textbf{(g)} Moment 1 residuals from Keplerian. 
\textbf{(h)} Filtered moment 1. 
\textbf{(i)} ALMA $^{13}$CO filtered moment 0. 
\textbf{(j)} ALMA $^{13}$CO filtered moment 2. 
\textbf{(k)} ALMA $^{13}$CO filtered moment 1. 
\textbf{(l)} ALMA C$^{18}$O filtered moment 1 (robust $1.5$). 
Panel (l) inset overlays the VLT/SPHERE $H$-band scattered light spirals S1-S7 (ref.\cite{hashimoto2011, fukagawa2004}) in red, and $^{13}$CO spirals S1-S9 we identify in black. 
}}
\label{extfig:kinematics-spirals}
\end{figure}

\begin{figure}[ht]
\includegraphics[width=\textwidth,angle=0]{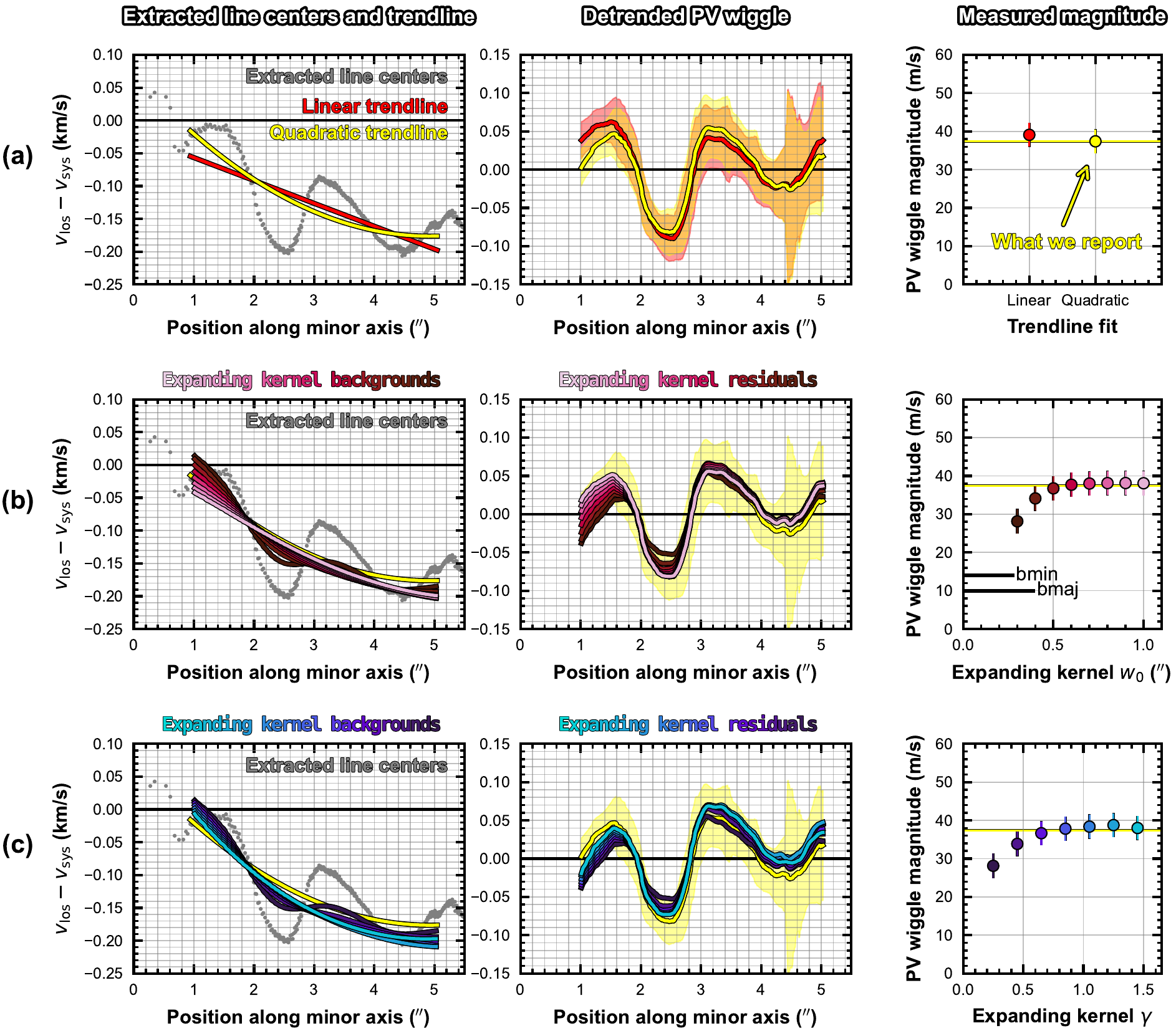}
\caption{{\bf 
\rev{Methods for isolating the sinusoidal component of the southern minor axis PV wiggle in the AB Aur disk.}} 
\rev{\textbf{(a)} Detrending the ALMA $^{13}$CO line centers from Figure~\ref{fig:3:spectra}a with linear and quadratic trendlines found by a least-squares fit. 
\textbf{(b)} Detrending with the expanding kernel high-pass filter, varying the kernel width parameter $w_0$ and keeping the kernel radial power-law index fixed to $\gamma=0.25$ (Equation \ref{eqn:expanding_kernel}). We find the background trendlines by extracting the velocity values from the high-pass filter background map (e.g. Extended Data Figure~\ref{extfig:hp-vkep-residuals}d) within the same $0.5^{\circ}$-wide wedge-shaped mask as we do for the line centers, positioned along the southern disk minor axis. 
\textbf{(c)} Like (b), but varying $\gamma$ and keeping $w_0$ fixed to $w_0=0.30''$. 
The high-pass filter detrending approach converges to the same measured PV wiggle magnitude as the quadratic fit approach.}
}
\label{extfig:comment-2D-convergence}
\end{figure}

\begin{figure}[ht]
\begin{center}
    \includegraphics[width=0.5\textwidth,angle=0]{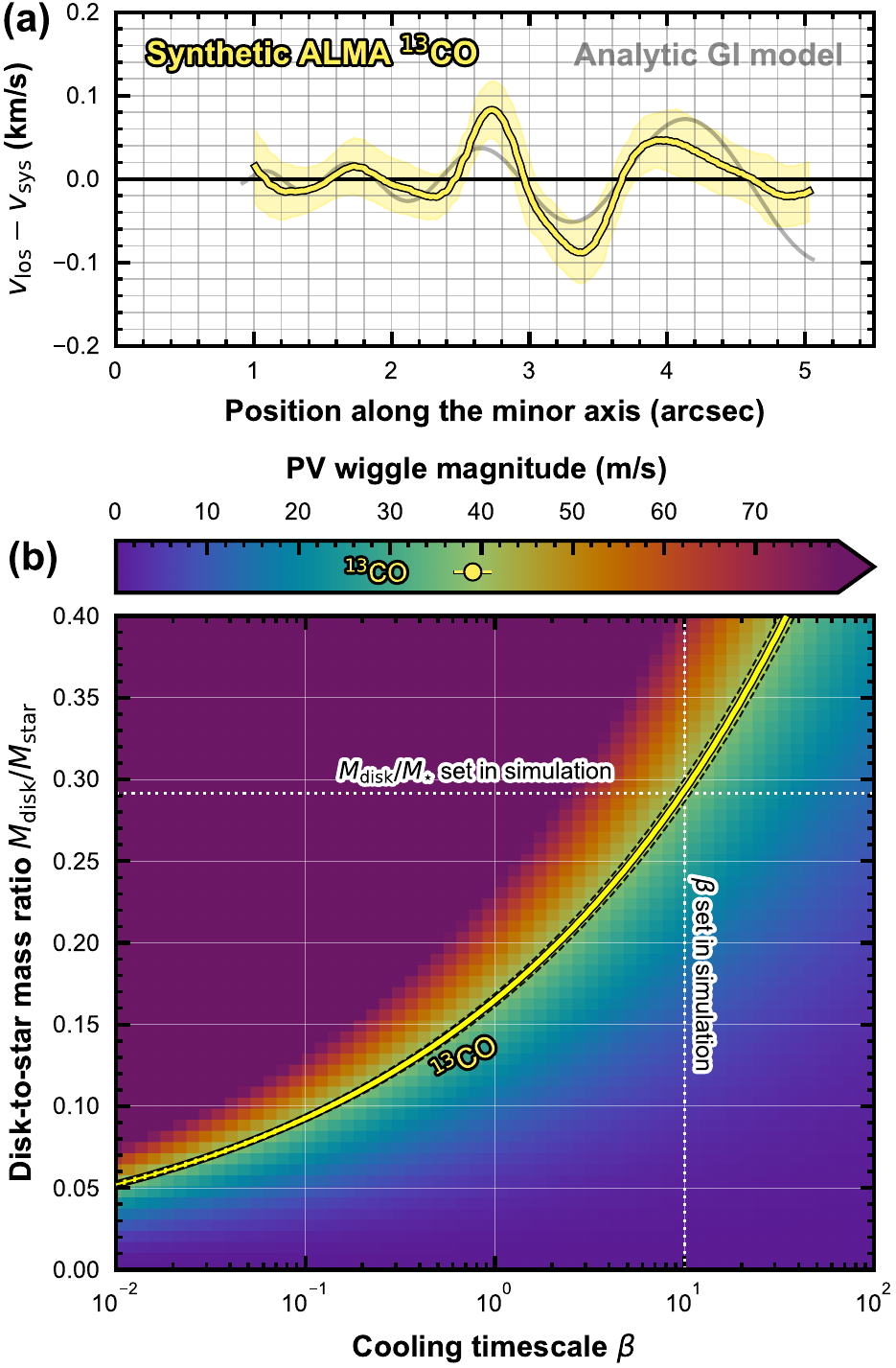}
\end{center}
\caption{\rev{{\bf PV wiggle morphology, magnitude, and disk mass recovery in the SPH GI disk simulation.}} 
Like Figure~\ref{fig:4:amplitude}, but for the synthetic ALMA observations of the SPH GI disk simulation. 
\textbf{(a)} The synthetic \rev{ALMA} $^{13}$CO line centers \rev{along the southern minor axis from Figure~\ref{fig:3:spectra}c, after quadratic detrending. Uncertainties on the line centers are shown by yellow shaded regions.} 
The magnitude of th\rev{is} PV wiggle is measured to be $\rev{39.1} \pm 1.9$ m/s. 
The analytic model shown in the background for qualitative comparison has the same parameters \rev{as the underlying SPH simulation} ($M_{\rm disk}/M_{\star}=0.29$ and $\beta=10$) and \rev{its} PV wiggle magnitude \rev{is} $39.0$ m/s. 
\textbf{(b)} \rev{A}s in Figure~\ref{fig:4:amplitude}c, \rev{a map of} the minor axis PV wiggle magnitude of $60\times60$ analytic models on a grid of disk-to-star mass ratios and cooling timescales.  
A contour is drawn at the measured magnitude \rev{of the synthetic $^{13}$CO PV wiggle in panel (a),} 
and dashed lines represent the quoted uncertainties. 
The technique successfully recovers the disk mass set in the SPH simulation.
}
\label{extfig:pv-wiggle-amp:sim}
\end{figure}

\begin{figure}[ht]
\includegraphics[width=\textwidth,angle=0]{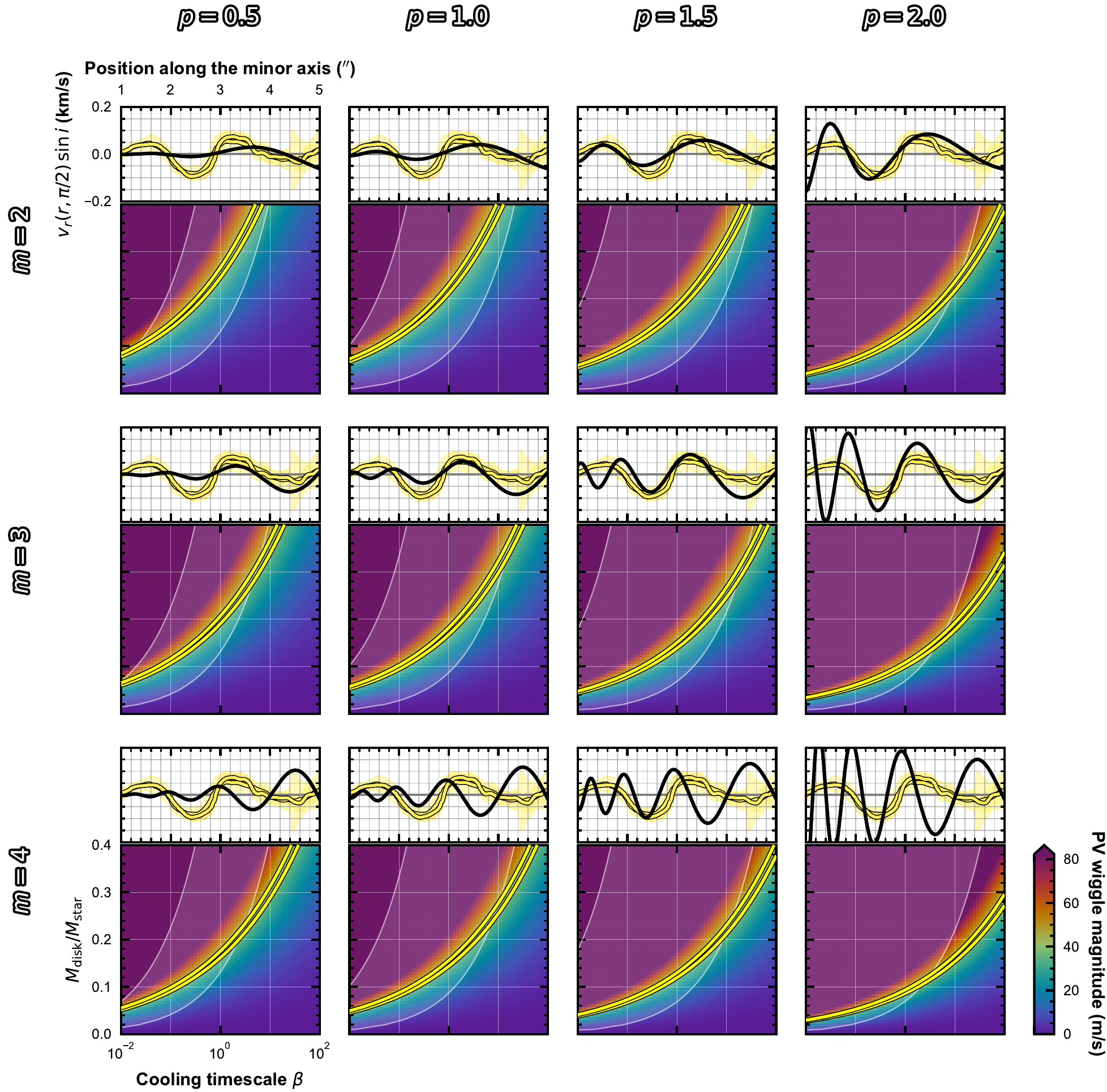}
\caption{{\bf \rev{Comparisons to additional sets of analytic models.}}  
\rev{Like Figure~\ref{fig:4:amplitude}, but varying the azimuthal wavenumber $m$ and surface density power-law index $p$ in the comparison grid of analytic GI model disks. Each upper subpanel shows the quadratically detrended $^{13}$CO and C$^{18}$O line centers (yellow) behind a demonstrative analytic PV wiggle (black) computed with the combination of $m$ and $p$ indicated by the row and column labels (keeping $M_{\rm disk}/M_{\star}=0.3$ and $\beta=10$ fixed). 
Each lower subpanel shows the corresponding map of PV wiggle magnitude computed for a $60\times60$ grid of analytic models in $M_{\rm disk}/M_{\star}$ and $\beta$, again with the combination of $m$ and $p$ indicated by the row and column labels. 
The two yellow contours are drawn at the magnitude values measured for the observed AB Aur $^{13}$CO and C$^{18}$O southern minor axis PV wiggles. 
\frev{The white shaded region between two white curves represents plausible $\beta$ ranges from $r=1-5''$.} 
The combination shown in Figure~\ref{fig:4:amplitude}c is $m=3$, $p=1.0$.
}}
\label{extfig:effect-p-m}
\end{figure}

\begin{figure}[ht]
\includegraphics[width=\textwidth,angle=0]{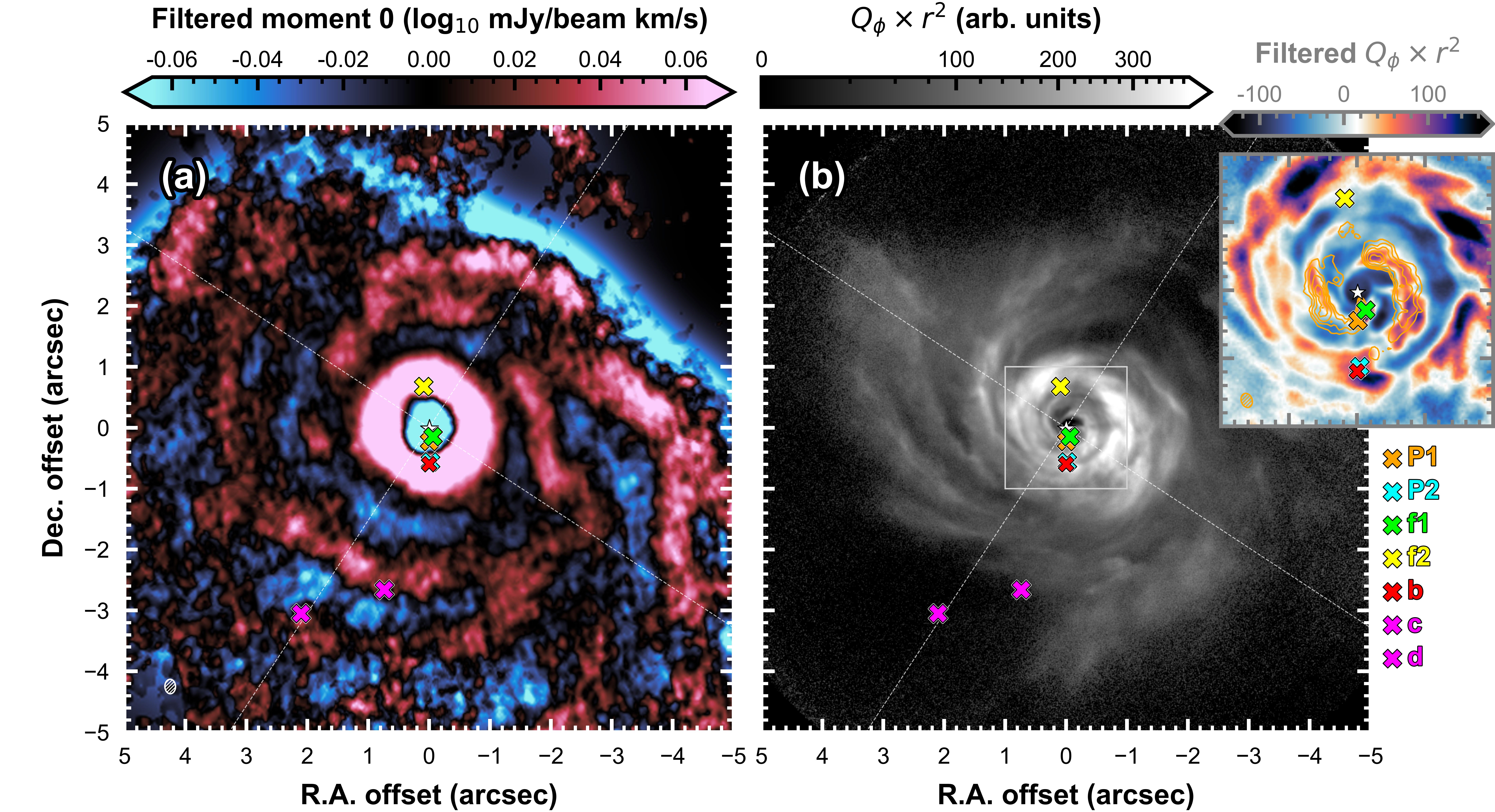}
\caption{{\bf Candidate sites of planet formation.} 
\rev{Coloured $\times$'s mark the locations of candidate protoplanets reported in the literature\cite{tang2017-abaur12COspirals, boccaletti2020-abaursphere, currie2022-abaurb}. 
A table providing the candidates' coordinates on the sky, estimated masses and the reporting references is available as \textbf{source data}.
\textbf{(a)} Filtered ALMA $^{13}$CO moment 0 map, as in Figure \ref{fig:1:residuals}c.
\textbf{(b)} VLT/SPHERE $H$-band scattered light image  (ref.\cite{boccaletti2020-abaursphere}), as in Figure \ref{fig:1:residuals}a.
The inset zooms into the central $2'' \times 2''$ region to show the spiral structures in different tracers at spatial scales unresolved by the present ALMA observations. 
The 
$H$-band scattered light image is shown after high-pass filtering, and  orange contours show the two spirals identified in ALMA $^{12}$CO $J=2-1$ moment 0 (ref.\cite{tang2017-abaur12COspirals}) 
at levels from $25$ to $50$ mJy/beam km/s in increments of $5$ mJy/beam km/s. 
} 
}
\label{extfig:candidate-sites-planets}
\end{figure}


\clearpage

\end{bibunit}


\end{document}